\documentclass[traditabstract]{aa}  
\usepackage{fixltx2e}
\usepackage[english]{babel}
\usepackage{graphicx,amsmath}
\usepackage{epstopdf}
\usepackage{epsf,color}
\usepackage[mathscr]{eucal}
\usepackage{amsmath}
\usepackage{amssymb,amsfonts}
\usepackage{natbib}
\usepackage{graphicx}
\usepackage{txfonts}
\usepackage{dsfont}
\definecolor{Mygreen}{rgb}{0.00, 0.72, 0.0}
\definecolor{Mypink}{rgb}{1.0, 0.0, 0.5}
\usepackage[breaklinks, citecolor=blue, linkcolor=Mygreen, urlcolor=Mypink, colorlinks=true, debug, baseurl=' ']{hyperref}
\usepackage{float}
\usepackage{color}
\def\simlt{\lower.5ex\hbox{$\; \buildrel < \over \sim \;$}}
\def\simgt{\lower.5ex\hbox{$\; \buildrel > \over \sim \;$}}

\bibpunct{(}{)}{;}{a}{}{,}
\newfont{\gwpfont}{cmssq8 scaled 1000}
\newcommand{\rexcess}{{\gwpfont REXCESS}}
\begin{document}

\def\aj{AJ}%
\def\araa{ARA\&A}%
\def\apj{ApJ}%
\def\apjl{ApJ}%
\def\apjs{ApJS}%
\def\aap{A\&A}%
 \def\aapr{A\&A~Rev.}%
\def\aaps{A\&AS}%
\def\mnras{MNRAS}
\def\ssr{SSRv}
\def\nat{Nature}
\def\jcap{JCAP}

\def\Mgv{M_{\rm g,500}}
\def\Mg{M_{\rm g}}
\def\YX {Y_{\rm X}}
\def\LXv {L_{\rm X,500}}
\def\TX {T_{\rm X}}
\def\fgv {f_{\rm g,500}}
\def\fg  {f_{\rm g}}
\def\kT {{\rm k}T}
\def\ne {n_{\rm e}}
\def\Mv {M_{\rm 500}}
\def \Rv {R_{500}}
\def\keV {\rm keV}
\def\Yv{Y_{500}}

\def\MT {$M$--$T_{\rm X}$}
\def\MYX {$M$--$Y_{\rm X}$}
\def\MMg {$M_{500}$--$M_{\rm g,500}$}
\def\MgT {$M_{\rm g,500}$--$T_{\rm X}$}
\def\MgY {$M_{\rm g,500}$--$Y_{\rm X}$}

\def\msol {{\rm M_{\odot}}}

\def\lesssim{\mathrel{\hbox{\rlap{\hbox{\lower4pt\hbox{$\sim$}}}\hbox{$<$}}}}
\def\gtrsim{\mathrel{\hbox{\rlap{\hbox{\lower4pt\hbox{$\sim$}}}\hbox{$>$}}}}

\def\psz{PSZ1\,G045.85$+$57.71}

% satellites
\def\xmm{XMM-{\it Newton}}
\def\planck{{\it Planck}} 
\def\chandra{{\it Chandra}}
\def \rosat {\hbox{\it ROSAT}}
\newcommand{\excpres}{{\gwpfont EXCPRES}}
\newcommand{\ma}[1]{\textcolor{red}{{ #1}}}
%###############################################################################################
%##########################              START THE PAPER             ##########################################
%###############################################################################################
\title{Non-parametric deprojection of NIKA SZ observations: Pressure distribution in the \planck-discovered cluster \mbox{PSZ1\,G045.85+57.71}}
\author{F.~Ruppin \inst{\ref{LPSC}}\thanks{Corresponding author: Florian Ruppin, \url{ruppin@lpsc.in2p3.fr}}
\and R.~Adam \inst{\ref{LPSC},\ref{OCA}}
\and  B.~Comis \inst{\ref{LPSC}}
\and  P.~Ade \inst{\ref{Cardiff}}
\and  P.~Andr\'e \inst{\ref{CEA}}
\and  M.~Arnaud \inst{\ref{CEA}}
\and  A.~Beelen \inst{\ref{IAS}}
\and  A.~Beno\^it \inst{\ref{Neel}}
\and  A.~Bideaud \inst{\ref{Neel}}
\and  N.~Billot \inst{\ref{IRAME}}
\and  O.~Bourrion \inst{\ref{LPSC}}
\and  M.~Calvo \inst{\ref{Neel}}
\and  A.~Catalano \inst{\ref{LPSC}}
\and  G.~Coiffard \inst{\ref{IRAMF}}
\and  A.~D'Addabbo \inst{\ref{Neel},\ref{Roma}}
\and  M.~De~Petris \inst{\ref{Roma}}
\and  F.-X.~D\'esert \inst{\ref{IPAG}}
\and  S.~Doyle \inst{\ref{Cardiff}}
\and  J.~Goupy \inst{\ref{Neel}}
\and  C.~Kramer \inst{\ref{IRAME}}
\and  S.~Leclercq \inst{\ref{IRAMF}}
\and  J.F.~Mac\'ias-P\'erez \inst{\ref{LPSC}}
\and  P.~Mauskopf \inst{\ref{Cardiff},\ref{Arizona}}
\and  F.~Mayet \inst{\ref{LPSC}}
\and  A.~Monfardini \inst{\ref{Neel}}
\and  F.~Pajot \inst{\ref{IAS}}
\and  E.~Pascale \inst{\ref{Cardiff}}
\and  L.~Perotto \inst{\ref{LPSC}}
\and  G.~Pisano \inst{\ref{Cardiff}}
\and  E.~Pointecouteau \inst{\ref{IRAP},\ref{IRAP2}}
\and  N.~Ponthieu \inst{\ref{IPAG}}
\and  G.W.~Pratt \inst{\ref{CEA}}
\and  V.~Rev\'eret \inst{\ref{CEA}}
\and  A.~Ritacco \inst{\ref{LPSC}}
\and  L.~Rodriguez \inst{\ref{CEA}}
\and  C.~Romero \inst{\ref{IRAMF}}
\and  K.~Schuster \inst{\ref{IRAMF}}
\and  A.~Sievers \inst{\ref{IRAME}}
\and  S.~Triqueneaux \inst{\ref{Neel}}
\and  C.~Tucker \inst{\ref{Cardiff}}
\and  R.~Zylka \inst{\ref{IRAMF}}}

\institute{
Laboratoire de Physique Subatomique et de Cosmologie, Universit\'e Grenoble Alpes, CNRS/IN2P3, 53, avenue des Martyrs, Grenoble, France
  \label{LPSC}
  \and
  Laboratoire Lagrange, Universit\'e C\^ote d'Azur, Observatoire de la C\^ote d'Azur, CNRS, Blvd de l'Observatoire, CS 34229, 06304 Nice cedex 4, France
  \label{OCA}
  \and
Astronomy Instrumentation Group, University of Cardiff, UK
  \label{Cardiff}
\and
Laboratoire AIM, CEA/IRFU, CNRS/INSU, Universit\'e Paris Diderot, CEA-Saclay, 91191 Gif-Sur-Yvette, France 
  \label{CEA}
\and
Institut d'Astrophysique Spatiale (IAS), CNRS and Universit\'e Paris Sud, Orsay, France
  \label{IAS}
\and
Institut N\'eel, CNRS and Universit\'e Grenoble Alpes, France
  \label{Neel}
\and
Institut de RadioAstronomie Millim\'etrique (IRAM), Granada, Spain
  \label{IRAME}
\and
Institut de RadioAstronomie Millim\'etrique (IRAM), Grenoble, France
  \label{IRAMF}
\and
Dipartimento di Fisica, Sapienza Universit\`a di Roma, Piazzale Aldo Moro 5, I-00185 Roma, Italy
  \label{Roma}
\and
Institut de Plan\'etologie et d'Astrophysique de Grenoble (IPAG), CNRS and Universit\'e Grenoble Alpes, France
  \label{IPAG}
    \and
School of Earth and Space Exploration and Department of Physics, Arizona State University, Tempe, AZ 85287
  \label{Arizona}
\and
Universit\'e de Toulouse, UPS-OMP, Institut de Recherche en Astrophysique et Plan\'etologie (IRAP), Toulouse, France
  \label{IRAP}
\and
CNRS, IRAP, 9 Av. colonel Roche, BP 44346, F-31028 Toulouse cedex 4, France 
  \label{IRAP2}
}

%\author{version 2.1}
%\date{Received \today \ / Accepted -- \ma{Version 7}}

\abstract {The determination of the thermodynamic properties of clusters of galaxies at intermediate and high redshift can bring new insights into the formation of large-scale structures. It is  essential for a robust calibration of the mass-observable scaling relations and their scatter,  which are key ingredients  for precise cosmology  using cluster statistics. Here we illustrate an application of high resolution  $(< 20$ arcsec) thermal Sunyaev-Zel'dovich (tSZ) observations by probing the intracluster medium (ICM) of the \planck-discovered galaxy cluster \psz\ at redshift $z = 0.61$, using tSZ data obtained with the NIKA camera, which is a  dual-band (150 and 260~GHz) instrument operated at the IRAM 30-meter telescope. We deproject jointly NIKA and \planck\ data to extract the electronic pressure distribution from the cluster core ($R \sim 0.02\, R_{500}$) to its outskirts ($R \sim 3\, R_{500}$) non-parametrically  for the first time at intermediate redshift. The constraints on the resulting pressure profile allow us to reduce the relative uncertainty on the integrated Compton parameter by a factor of two compared to the \planck\ value. Combining the tSZ data and the deprojected electronic density profile from \xmm\ allows us to undertake a hydrostatic mass analysis, for which we study the impact of a spherical model assumption on the total mass estimate. We also investigate the radial temperature and entropy distributions. These data indicate that \psz\ is a massive ($M_{500} \sim 5.5 \times 10^{14}$ M$_{\odot}$) cool-core cluster. This work is part of a pilot study aiming at optimizing the treatment of the NIKA2 tSZ large program dedicated to the follow-up of SZ-discovered clusters at intermediate and high redshifts. This study illustrates the potential of NIKA2 to put constraints on the thermodynamic properties and tSZ-scaling relations of these clusters, and demonstrates the excellent synergy between tSZ and X-ray observations of similar angular resolution.}

\titlerunning{Non parametric deprojection of NIKA SZ observations}
\authorrunning{F. Ruppin \emph{et al.}}
\keywords{Instrumentation: high angular resolution -- Galaxies: clusters: individual: \mbox{PSZ1\,G045.85+57.71}; intracluster medium}
\maketitle

%#############################################################################################
%##########################                             INTRODUCTION                               ##########################%#############################################################################################
\section{Introduction}\label{sec:Introduction}
%---------- Galaxy cluster cosmology 
Galaxy clusters are the ultimate manifestation of the hierarchical structure formation process in the standard cosmological model, and as such, they are sensitive to both the matter content and expansion history of the Universe in which they form. Clusters are thus potentially powerful tools to infer cosmological parameters. In particular, counting clusters as a function of their mass and redshift \citep[\emph{e.g.,}][]{seh11,Planck_SZ_cosmo2015,deh16} brings constraints on the cosmological parameters that are complementary to those derived with other probes such as type Ia supernovae \citep[\emph{e.g.,}][]{rie07}, the CMB temperature and polarization angular power spectra \citep[\emph{e.g.,}][]{Planck_param}, or baryonic acoustic oscillations \citep[\emph{e.g.,}][]{and14}.\\
%---------- Galaxy cluster content
About 85\% of the total mass in galaxy clusters is from dark matter. The principal baryonic component is found in the hot, ionized, X-ray emitting gas of the intracluster medium (ICM), containing about 12\% of the total mass. The remaining baryonic mass is found in the stellar population.  Cluster masses can be inferred from several independent observables.
%---------- Galaxy cluster mass estimation
The velocity dispersion of the galaxies \citep[\emph{e.g.,}][]{biv06,sif16}, various X-ray properties such as temperature or luminosity \citep[\emph{e.g.,}][]{SVM_prof,pra09}, or the lensing distortions of background galaxies \citep[\emph{e.g.,}][]{app14,ume14,hoe15}  can be related to the underlying total mass. Another observational probe of interest is the thermal Sunyaev-Zel'dovich effect \citep[tSZ;][]{sun72}, which is due to the inverse Compton scattering of cosmic microwave background (CMB) photons with high-energy electrons of the ICM. As this effect is directly proportional to the thermal energy contained in the ICM, it is expected to provide a low scatter mass proxy for galaxy clusters \citep[\emph{e.g.,}][]{das04,nag07}. Furthermore, as the tSZ effect  is a CMB spectral distortion, it does not suffer from cosmological dimming. This observable is therefore a powerful probe to estimate both galaxy cluster total mass and baryonic content distribution up to high redshift.\\
%---------- Need for high-z, high resolution and multi-probe analysis
The \planck\ satellite, the South Pole Telescope (SPT), and the Atacama Cosmology Telescope (ACT) surveys have used tSZ observations to discover and characterize large galaxy cluster samples \citep[\emph{e.g.,}][]{Planck_cata2,SPTcluster,ACT_cluster}. In addition, individual observations of known clusters have been obtained with a number of instruments, such as APEX-SZ, CARMA, SZA, BOLOCAM, and AMIs \citep[\emph{e.g.,}][]{Apexsz,Carmasz,SZA,Bolocamsz,AMI_followup}. However, their relatively low angular resolution ($> 1$ arcmin) restricts the tSZ characterization of the ICM to low redshift \citep{pla10, NonparamPressure,bon12,Planck_pressure_prof,SayersPointSource}, as a combination with higher resolution X--ray observations is needed to map clusters at both large and small scales \citep[\emph{e.g.,}][]{Planck_pressure_prof,eck13}.\\
In combination with local data, high angular resolution tSZ observations at intermediate to high-redshift ($z > 0.5$) have a number of different applications. They can be used to study the evolution of structural properties such as cluster pressure profiles and their scatter. Furthermore, they provide new insights and constraints on scaling properties such as the relation between the integrated Compton parameter and the cluster total mass and its scatter. High angular resolution tSZ observations can also be used to characterize the two-dimensional (2D) pressure distribution within the ICM. This information is essential for understanding cluster formation physics and performing precise cosmological analysis with the cluster population.\\
Cluster growth and evolution is characterized by complex astrophysical phenomena, including deviation from equilibrium and generation of turbulence due to merging events and feedback from active galactic nuclei. While stochastic, the frequency of these events evolves with time and increases at high redshift. They are the prime cause of scatter and deviations from self-similarity in the scaling relations that are used to link observables to mass in cosmological analyses (\emph{e.g.,} \citealt{mergers_bias,MUSIC_scaling}). Of particular importance is the clarification of the physical origin of this normalization and scatter in the scaling relations, rendering the use of galaxy clusters for cosmological application more robust.\\
%---------- NIKA
The New IRAM KIDs Array \citep[NIKA;][]{NIKA_cam1,NIKA_elec2,cal13} was a dual-band continuum camera operated at the Institut de Radio Astronomie Millimetrique (IRAM) 30 m telescope between 2010 and 2015. It was one of the very few tSZ instruments with sub-arcminute resolution. Other examples include the Goddard-IRAM Superconducting 2-Millimeter Observer \citep[GISMO;][]{sta08} and the Multiplexed SQUID TES array at Ninety Gigahertz \citep[MUSTANG;][]{kor11}.  NIKA was the only dual-band sub-arcminute instrument \citep{NIKA_calib} that observed
the tSZ effect simultaneously at 150 and 260~GHz with an angular resolution
of 18.2 and 12.0~arcsec, respectively. Furthering  the characterization of galaxy cluster pressure profiles that has been initiated by arcminute resolution instruments at low redshift, NIKA  has now mapped the pressure distribution in a number of galaxy clusters at intermediate and high redshift \citep[see][]{RXJ1347NIKA,CLJ1227NIKA,MACSJ1424NIKA,MACSJ0717NIKA}.\\
%---------- Main ideas developped in the paper
In this paper we detail the NIKA observations of the \planck-discovered cluster \psz\ at $z=0.61$. A key result is the first non-parametric measurement with high statistical precision of the pressure profile of a distant cluster at an angular resolution $\sim 20$ arcsec, extending to much higher redshift pioneering non-parametric pressure profile measurements at low resolution \citep{NonparamPressure}. \citealt{NonparamPressure} have applied the deprojection method presented in \citealt{NonparamNord} to the APEX-SZ data \citep{NonparamApex} of the nearby cluster Abell 2204 ($z=0.15$). They have shown that a non-parametric modeling of the gas pressure profile can be obtained. Previous works have shown that deprojection methods can be used to probe the ICM of clusters from simulations \citep{NonparamSimu1,NonparamSimu2,NonparamSimu3}.\\
The work presented in this paper is a pilot study for the forthcoming SZ observations (see \citealt{ProceedingMoriond}) with NIKA2 (see \citealt{ProceedingSPIE}). The combination with \planck\ data allows the determination of the non-parametric pressure profile out to scales of $\gtrsim 3$ Mpc, substantially improving the constraints on the spherically integrated Compton parameter. Using the deprojected gas density profile from \xmm, we reconstruct the thermodynamic properties of the ICM without making use of  X-ray spectroscopic information. This result illustrates the excellent synergy between tSZ and X-ray observations of similar angular resolution, and serves as a pilot study for combining tSZ data to measure the gas pressure with short X-ray observations to measure the gas density.\\
%---------- Paper organization
This paper is organized as follows. The NIKA observations of \psz\ at the IRAM 30-meter telescope and the raw data processing are explained in Sect. 2. Ancillary data, previous tSZ observations, point source contamination, and XMM-Newton data reduction, are described in Sect. 3. The modelization of the ICM and the method to estimate the cluster total mass are presented in Sect. 4. We also discuss the characterization of the cluster ellipticity and its impact on the mass estimation. In Sect. 5 a non-parametric multiprobe analysis is performed to extract the radial pressure profile and obtain the ICM thermodynamic properties. The conclusions and NIKA2 perspectives are discussed in Sect. 6. Throughout this study we assume a flat $\Lambda$CDM cosmology following the latest \planck\ results \citep{Planck_param}: H$_0 = 67.8$ km s$^{-1}$ Mpc$^{-1}$, $\Omega_{\rm m} = 0.308$, and $\Omega_\Lambda = 0.692$. Within this framework, at the cluster redshift, one arcsec corresponds to 6.93 kpc.

\begin{figure*}[h]
\centering
\includegraphics[height=6.8cm]{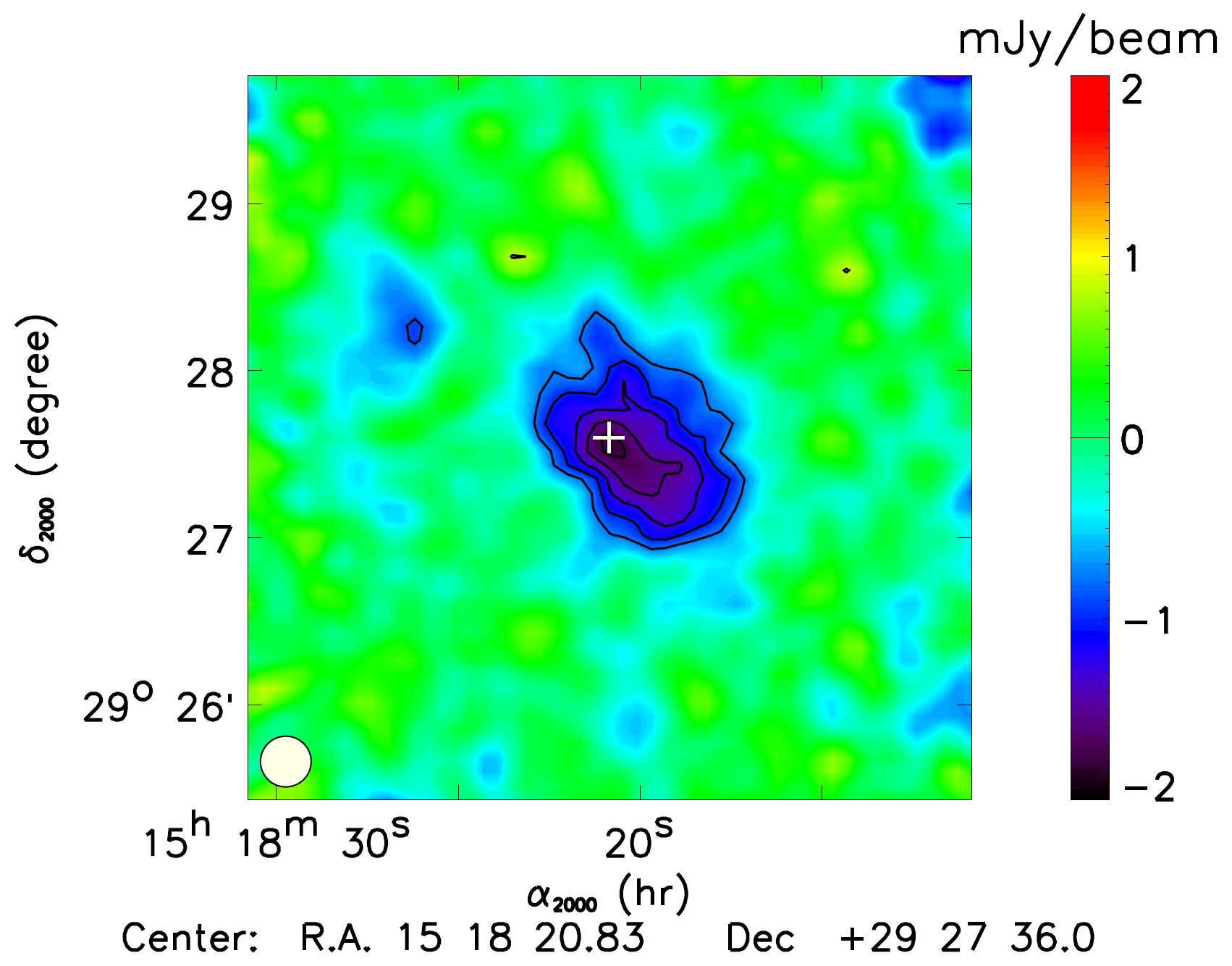}
\includegraphics[height=6.8cm]{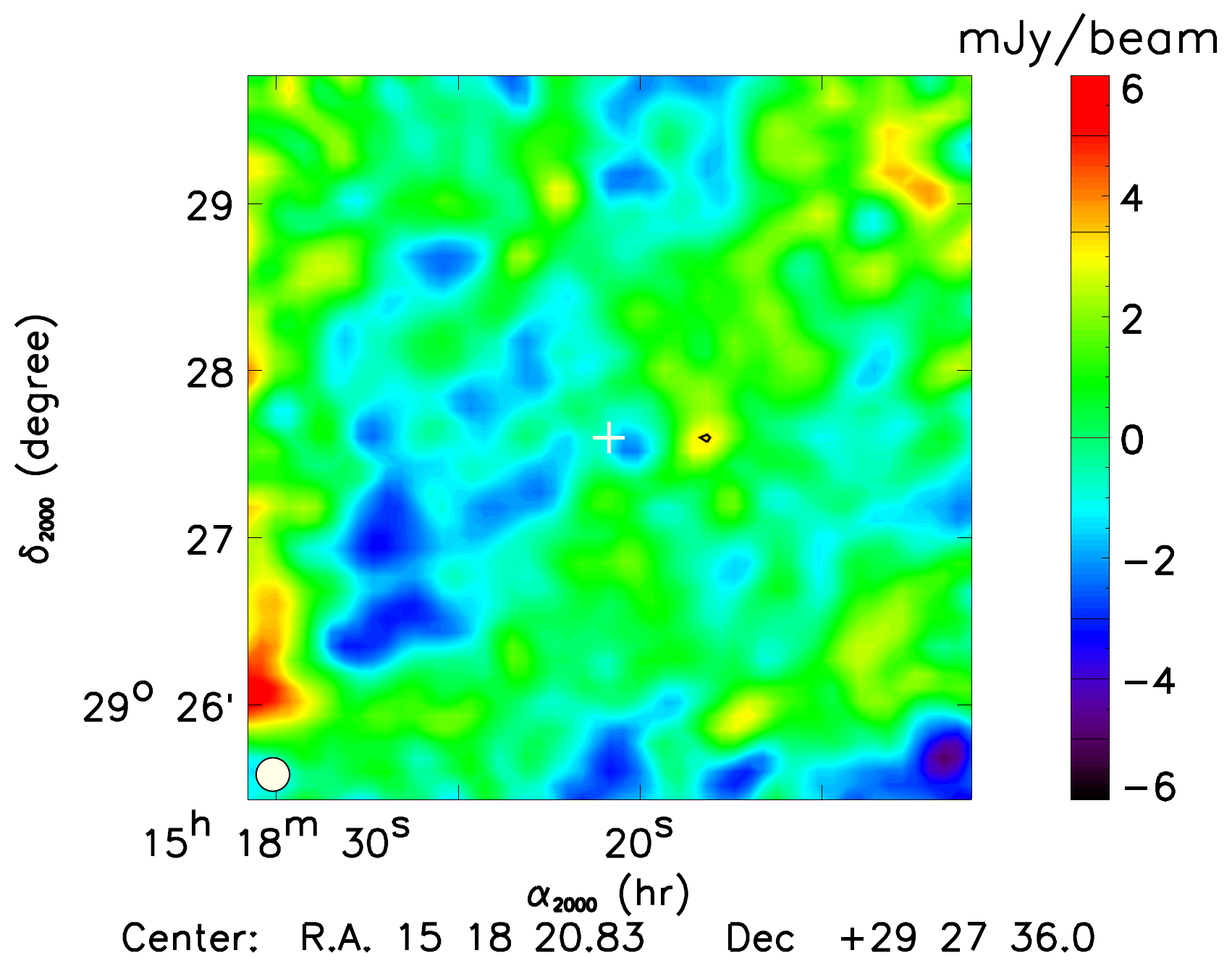}
\caption{{\footnotesize NIKA tSZ surface brightness maps at 150 GHz (left) and 260 GHz (right). The significance of the measured signal is given by the black contours starting at $3\sigma$ with $1\sigma$ spacing. The maps are smoothed with an additional 10~arcsec Gaussian filter for display purposes and the NIKA beam FWHMs are represented as  white disks in the bottom left-hand corner of the maps. The white crosses indicate the X-ray center. Note that we use the original maps (without additional smoothing) in the following analysis.}}
\label{fig:brightness_map}
\end{figure*}
%##########################################################################################
%##########################                         NIKA Observations                        ##########################%##########################################################################################
\section{Observation at the IRAM 30-meter telescope with NIKA}\label{sec:Observations}

We present in this section the NIKA observations of the \mbox{PSZ1\,G045.85+57.71} cluster, performed in October 2014, which have been used to produce the tSZ surface brightness maps at 150 GHz and 260 GHz. To begin with, we describe  the key elements of the thermal SZ effect. 

%========== tSZ
\subsection{The thermal Sunyaev-Zel'dovich effect}
The thermal Sunyaev-Zel'dovich effect corresponds to the Compton scattering of CMB photons on high-energy ICM electrons. The scattering equation describing this interaction was derived by Kompaneets (\citealt{Kompaneets}) in the case where the radiation temperature is negligible compared to the plasma temperature. Using this equation, Sunyaev and Zel'dovich (\citealt{sun72}, \citealt{SZ_effect}) computed the induced variation of the spectral radiance compared to the CMB variation as follows:
\begin{equation}
        \frac{\Delta I_{tSZ}}{I_0} = y \, f(\nu, T_e),
\label{eq:deltaI}
\end{equation}
where $y$ is the Compton parameter that describes the amplitude of the spectral distortion, $f(\nu, T_e)$ gives the frequency dependence of the tSZ spectrum, and $T_e$ is the ICM electronic temperature. The Compton parameter is related to the line-of-sight integral of the electronic pressure $P_e$,
\begin{equation}
        y = \frac{\sigma_{\mathrm{T}}}{m_{e} c^2} \int P_{e} \, dl,
        \label{eq:y_compton}
\end{equation}
where $m_{e}$ is the electron mass, $c$ the speed of light, and $\sigma_{\mathrm{T}}$ the Thomson scattering cross section. The integrated Compton parameter $Y_{tot}$ is then computed via the aperture photometry performed on the Compton parameter map\footnote{This definition gives the cylindrical Compton parameter of the cluster up to $5R_{500}$}.\\The frequency dependence of the tSZ spectrum is given by the expression (\citealt{Birkinshaw}, \citealt{Carlstrom})
\begin{equation}
        f(x, T_e) = \frac{x^4e^x}{(e^x-1)^2}\left(x \, \mathrm{coth}\left(\frac{x}{2}\right) - 4 \right) (1+\delta_{tSZ}(x, T_e)),
        \label{eq:tSZ_spectrum}
\end{equation}
$$\mathrm{with}~~~x = \frac{h\nu}{k_B T_{\mathrm{CMB}}},$$
where $h$ and $k_B$ are the Planck and Boltzmann constants, respectively, and $\delta_{tSZ}(x, T_e)$ corresponds to the relativistic correction, which is non-negligible for plasma temperatures larger than 10~keV (\citealt{relat_corr}). We thus notice that the spectral deformation induced by the tSZ effect is completely characterized by the $f$ function and does not depend on the plasma temperature if relativistic corrections are negligible. In this case, $f$ is positive (negative) for frequencies higher (lower) than 217 GHz. We therefore expect a negative signal on the 150 GHz NIKA map and a positive signal at 260 GHz. 

%========== Observations
\subsection{Observing conditions, scanning strategy, calibration, and data reduction}
%---------- Observations
\mbox{PSZ1\,G045.85+57.71} was observed by the NIKA camera simultaneously at 150 GHz and 260 GHz during the second NIKA open pool in November 2014. In this section we present the observation conditions, scanning strategy, calibration procedure, and data reduction method.\\
\indent The pointing center was chosen to be at (R.A., Dec. 2000) = (15:18:20.8, +29:27:36.75) following the \planck\ and XMM-{\it Newton} observations. All the coordinates in this paper are given in the equinox 2000 system. The mean zenith opacities were measured to be 0.21 and 0.27 at 150 and 260 GHz, respectively,  and the atmosphere was particularly unstable because of the presence of wind, which induces an increased residual noise on the final map (see \citealt{NIKA_calib} for details on the opacity measurement procedure with NIKA). The mean elevation of the source was 49 degrees. The effective number of valid detectors was 113 at 150 GHz and 156 at 260 GHz for a field of view of 1.9 and 1.8 arcmin, respectively.\\
%---------- The scanning strategy 
\indent The cluster was mapped using on-the-fly raster scans made by a succession of 19 subscans of 6 arcmin length at constant azimuth or elevation with 10 arcsec steps between subscans. After discarding data affected by high atmospheric instabilities, the overall effective observing time on the cluster is 4.35 hours.\\
%---------- Calibration
\indent We used Uranus as a primary calibrator and the Moreno model (\citealt{model_moreno}) to estimate its brightness temperature frequency dependence (see \citealt{RXJ1347NIKA}, \citealt{these_remi} and \citealt{NIKA_calib} for details on the calibration procedure). The primary beam was modeled by a Gaussian function with a FWHM that has been measured to be 18.2 and 12.0 arcsec at 150 and 260 GHz, respectively. Using the dispersion of the measured Uranus fluxes and the uncertainty on the Moreno model (accurate to 5\%; see \citealt{Planck_calib}), the overall calibration uncertainty is estimated to be 9 and 11\% at 150 and 260 GHz, respectively. We estimated the conversion factors from the measured surface brightness to the Compton parameter taking the NIKA bandpass measurements into account. We found the computed values to be $-11.1\pm 1.0$ and $3.4\pm 0.4$ Jy/beam per unit of Compton parameter accounting for calibration uncertainties at 150 and 260 GHz, respectively. The main instrumental characteristics of the NIKA camera during the second open pool are summarized in table \ref{tab:NIKA_instru}.\\
%---------- Reduction
\indent We follow the raw data reduction method detailed in (\citealt{CLJ1227NIKA}). The main steps of the procedure are briefly summarized here. 
\begin{table}[t]
\begin{center}
\begin{tabular}{ccc}
\hline
\hline
Observing band & 150 GHz & 260 GHz \\
\hline
Gaussian beam model FWHM (arcsec) & 18.2 & 12.0 \\
Field of view (arcmin) & 1.9 & 1.8 \\
Effective number of detectors & 113 & 156 \\
Sensitivity (mJy/beam $\rm{s^{1/2}}$) & 12 & 61 \\
Conversion factor $y$-Jy/beam & $-11.1\pm 1.0$ & $3.4\pm 0.4$ \\
Pointing errors (arcsec) & $<3$ & $<3$ \\
Calibration uncertainties & 9\% & 11\% \\
\hline
\hline
\end{tabular}
\end{center}
\caption{{\footnotesize Instrumental characteristics of NIKA at the IRAM 30-m telescope in November 2014.}}
\label{tab:NIKA_instru}
\end{table}
The selection of valid detectors is based on their noise properties and optical responses. We removed glitches in the timelines due to impacts of cosmic rays prior to the analysis. We suppressed fluctuations associated with cryogenic vibrations  in the Fourier domain. We removed the atmospheric and electronic correlated noise by subtracting the common-mode templates estimated by averaging the timelines for each array. We flagged the cluster using the S/N map estimation in an iterative way to avoid ringing and reduce the signal filtering. We estimated the resulting transfer function of the data processing using simulations and this function is fairly constant with a filtering of $\sim 5\%$ at scales smaller than the NIKA field of view but larger than the beam size. The filtering increases rapidly for larger scales (see Fig. 3 in \citealt{CLJ1227NIKA} as a typical example of the transfer function for this analysis). For each scan, the processed time order information is projected on a pixelized grid for the two NIKA frequencies. The computed scans are eventually coadded using inverse variance weighting to obtain the final maps shown in Fig. \ref{fig:brightness_map}.
\begin{figure*}[t]
\centering
\includegraphics[height=5.5cm]{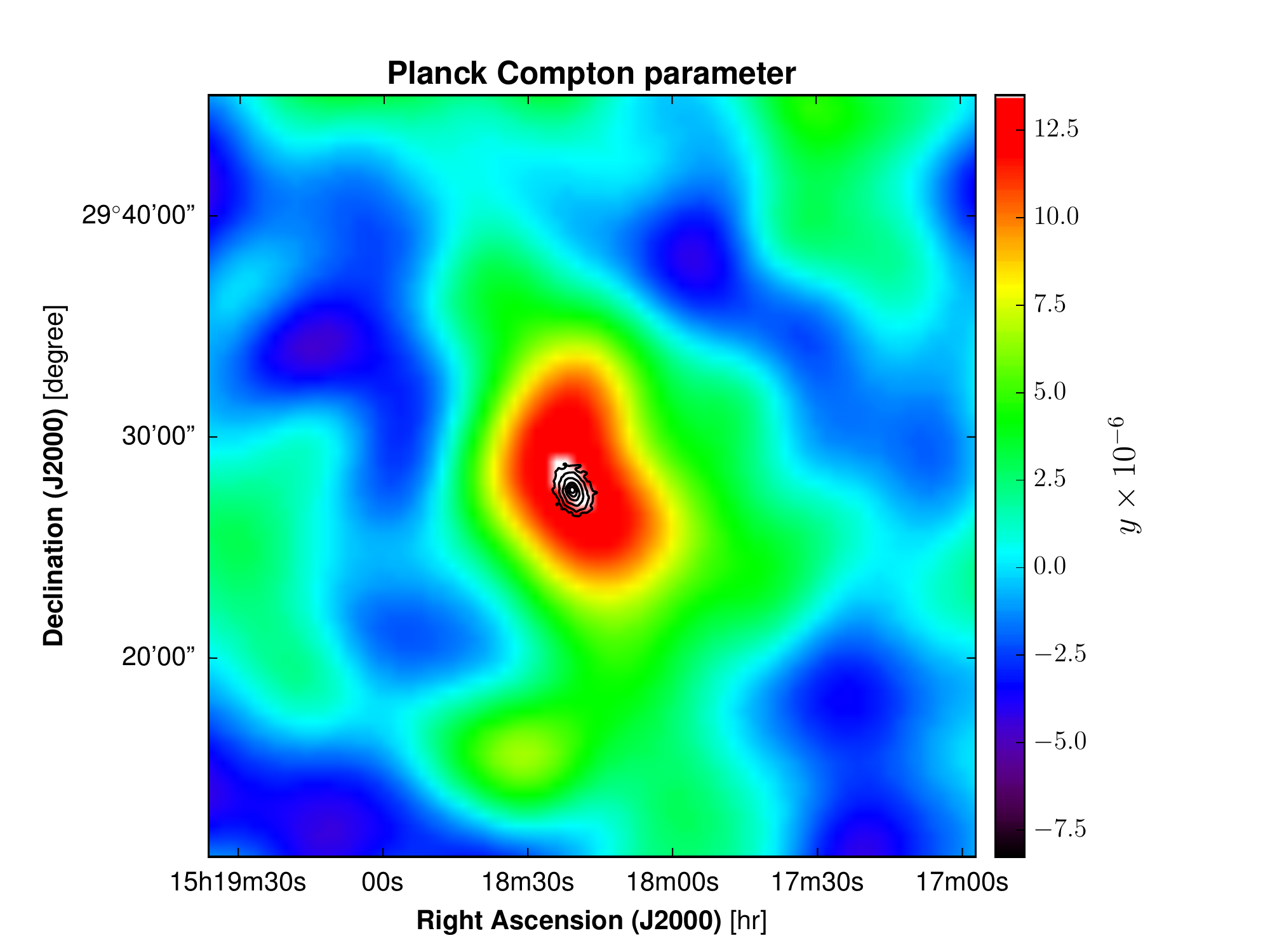}
\hspace{0.5cm}
\includegraphics[height=5.5cm]{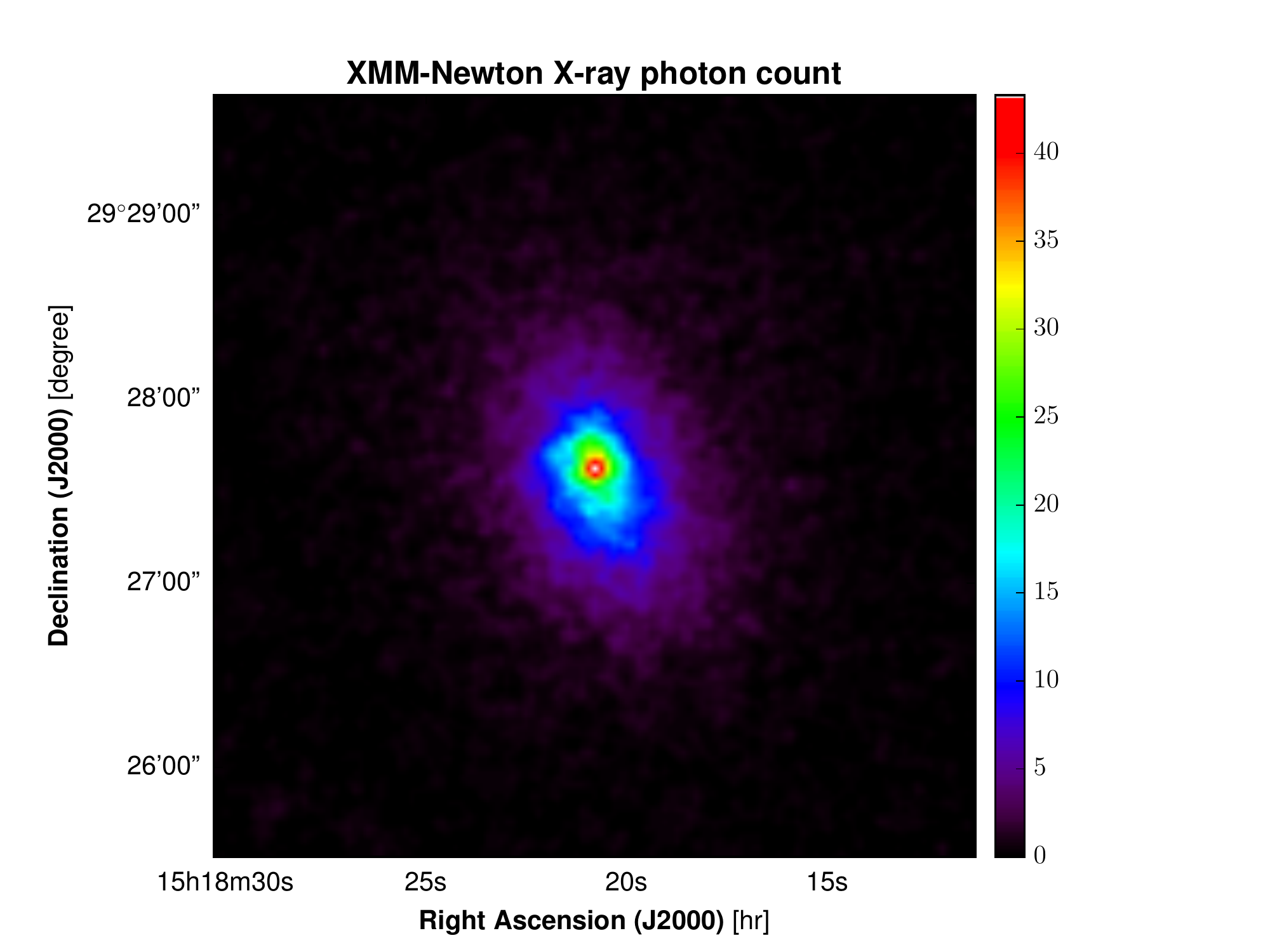}
\caption{{\footnotesize \textbf{Left:} \planck\ MILCA Compton parameter map of \mbox{PSZ1\,G045.85+57.71} obtained by extracting a patch of the \planck\ full sky $y$-map (\citealt{ymap_planck}). The XMM-{\it Newton} X-ray contours are overlaid in black to show the effect of the \planck\ beam dilution on the cluster inner structure. \textbf{Right:} XMM-{\it Newton} X-ray photon count map of \mbox{PSZ1\,G045.85+57.71} smoothed with an additional 4~arcsec Gaussian filter for display purposes.}}
\label{fig:Planck_XMM}
\end{figure*}
\begin{table*}[h]
\begin{center}
\begin{tabular}{ccccc}
\hline
\hline
Source & Identifier & Position & 1.4 GHz & Reference\\
 &  &  & [mJy] & \\
\hline
RS1 & FIRST J151819.5+292712 & 15h18m19.5s +29d27m13s  & $1.71 \pm 0.14$ & FIRST, \citealt{Becker1995} \\
RS2 & FIRST J151822.4+292917 & 15h18m22.5s +29d29m18s  & $2.9 \pm 0.5$  &  FIRST, \citealt{Becker1995} \\  
\hline
\hline
\end{tabular}
\end{center}
\caption{{\footnotesize Location and flux of the radio sources observed in the $4.4 \times 4.4$ arcmin$^2$ field around \mbox{PSZ1\,G045.85+57.71}.}}
\label{tab:Radio_ps}
\end{table*}
\begin{table*}[h]
\begin{center}
\begin{tabular}{ccccccccc}
\hline
\hline
Source & R.A. offset & Dec. offset & $F_{\mathrm{1~GHz}}$ & $\alpha_{\mathrm{radio}}$ & $F_{\mathrm{150~GHz}}$ & $F_{\mathrm{260~GHz}}$ & $\mathrm{RMS_{150~GHz}}$ & $\mathrm{RMS_{260~GHz}}$\\
 & [arcsec] & [arcsec] & [mJy] &  & [mJy] & [mJy] & [mJy] & [mJy]\\
\hline
RS1 & -19.4 & -23.9 & $2.2 \pm 0.2$ & $-0.7 \pm 0.2$ & $0.11 \pm 0.11$ & $0.075 \pm 0.092$ & 0.36 & 1.9\\
RS2 & 25.2 & 124.8 & $3.7 \pm 0.7$ & $-0.7 \pm 0.2$ & $0.18 \pm 0.19$ & $0.13 \pm 0.16$ & 0.42 & 2.2\\
\hline
\hline
\end{tabular}
\end{center}
\caption{{\footnotesize Best-fit parameters and extrapolation of the fluxes in the NIKA bands of the radio sources in the $4.4 \times 4.4 ~\rm{arcmin^2}$ field around \mbox{PSZ1\,G045.85+57.71}. The mean RMS noise on the flux of the identified point sources at their respective locations is also given at both NIKA frequencies. See text for details.}}
\label{tab:Radio_ps_flux}
\end{table*}
%========== NIKA maps
\subsection{NIKA observations}\label{sec:Raw_NIKA_observations}
%---------- The map
The NIKA tSZ surface brightness maps of \mbox{PSZ1\,G045.85+57.71} at 150 and 260~GHz are shown in figure \ref{fig:brightness_map}. The NIKA maps are centered on the X-ray peak coordinates denoted as a white cross and were smoothed with a 10 arcsec Gaussian filter for display purposes. We observe a strong tSZ decrement on the 150 GHz map, which reaches a $7\sigma$ significance at the surface brightness peak (-1.9 mJy/beam). The observed galaxy cluster exhibits an elliptical morphology with a major axis going from the southwest to the northeast of the center and does not indicate the presence of ICM substructure. Furthermore, the maximum tSZ decrement is aligned with the X-ray peak and thus does not indicate that \mbox{PSZ1\,G045.85+57.71} has a disturbed core. As expected, there is no significant tSZ signal on the 260~GHz map. Indeed, we can estimate the expected tSZ surface brightness peak at 260~GHz knowing the tSZ surface brightness at 150~GHz and the tSZ spectrum analytic expression (eq:\ref{eq:tSZ_spectrum}). The estimated value of \mbox{$\sim 1$~mJy/beam} is below the standard deviation of the residual noise in the NIKA 260~GHz map. Furthermore, the 260~GHz map does not present any significant submillimeter point source contamination given the RMS noise on this map.\\
%---------- The error bars
The residual noise on the map has to be characterized to estimate the significance contours of the measured signal at both NIKA frequencies. Furthermore, the noise covariance matrix $C_{\mathrm{NIKA}}$ at 150~GHz has to be estimated to be used for the ICM characterization.\\
\indent Following the procedure described in \citealt{MACSJ1424NIKA}, we use null-map realizations at 150 and 260~GHz to estimate the best-fit noise power spectrum models at both NIKA frequencies. The estimated residual noise power spectrum models, together with the integration time per pixel at 150 and 260~GHz, enable the simulation of Monte Carlo realizations of noise maps that are used to estimate the S/N on the final maps (see Fig. \ref{fig:brightness_map}) and to compute the noise covariance matrix at 150~GHz.

%#############################################################################################
%##########################                                Ancillary data                              ##########################%#############################################################################################

\section{\mbox{PSZ1\,G045.85+57.71} ancillary data}\label{sec:Previous}

This section presents the ancillary data obtained from previous observations of \mbox{PSZ1\,G045.85+57.71}. These data are used in the following multiprobe ICM characterization and give complementary information on the dynamical state of this cluster.
\begin{figure*}[t]
\centering
\includegraphics[height=3.9cm]{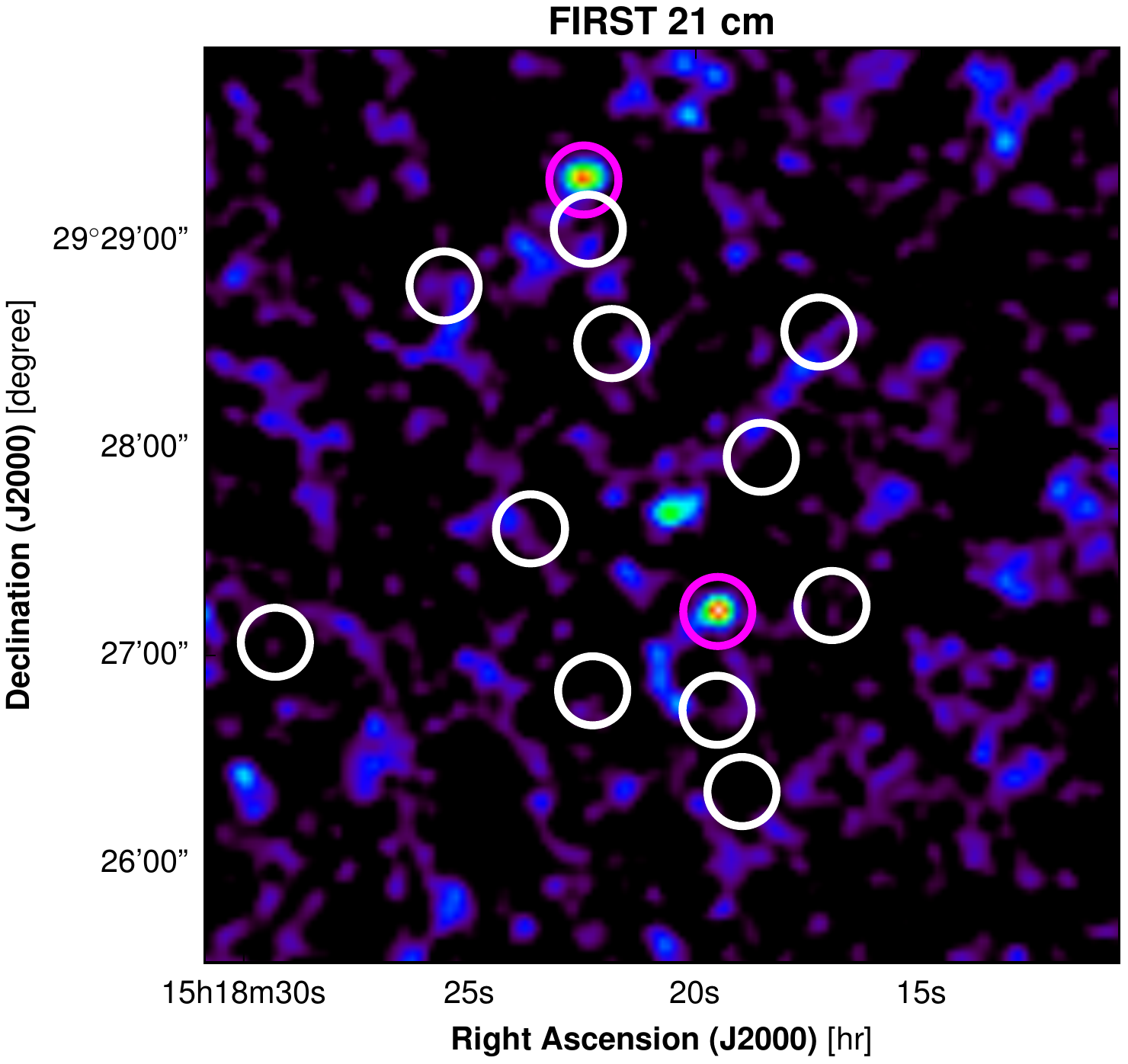}
\hspace{0.4cm}
\includegraphics[height=3.9cm]{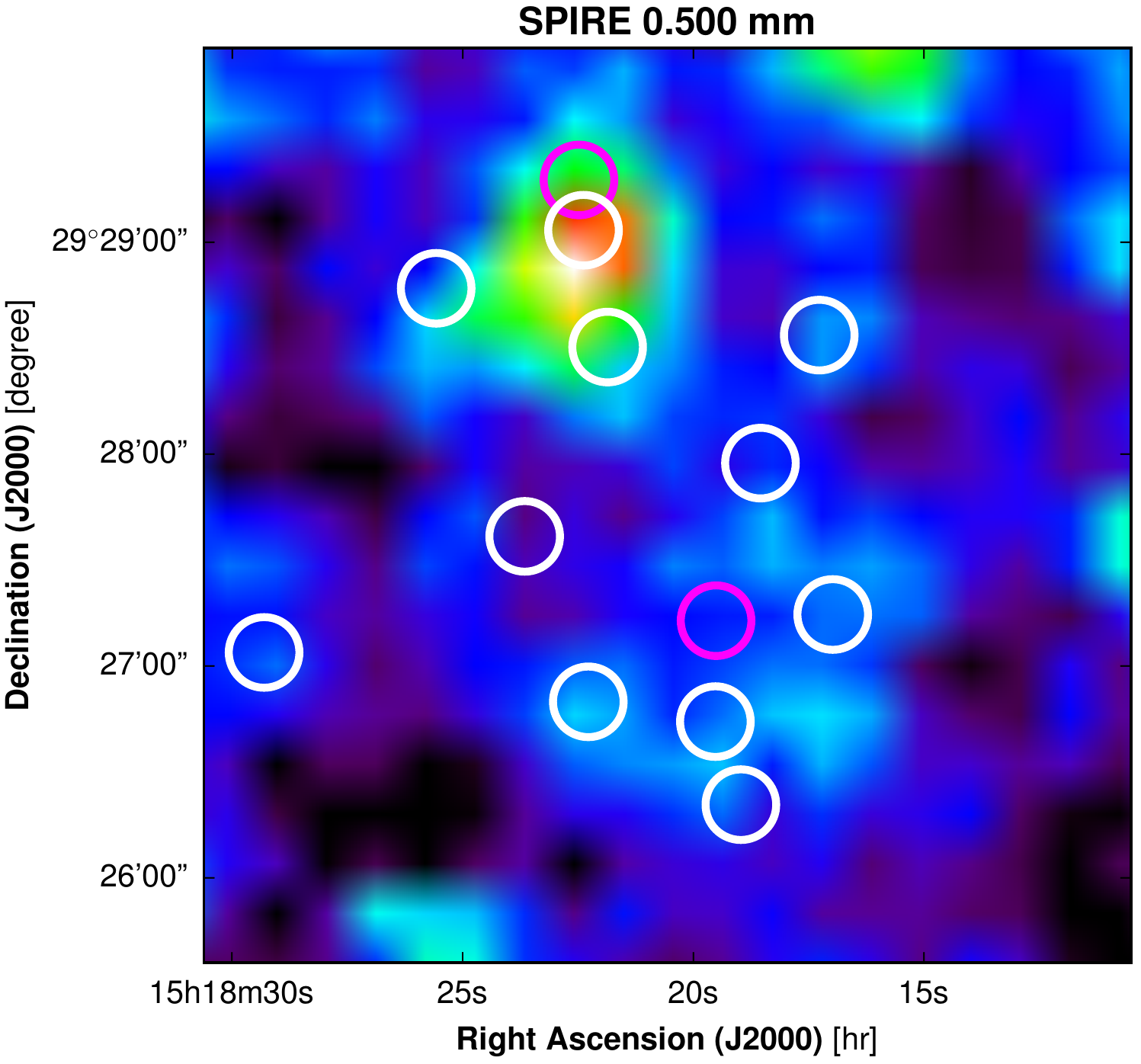}
\hspace{0.4cm}
\includegraphics[height=3.9cm]{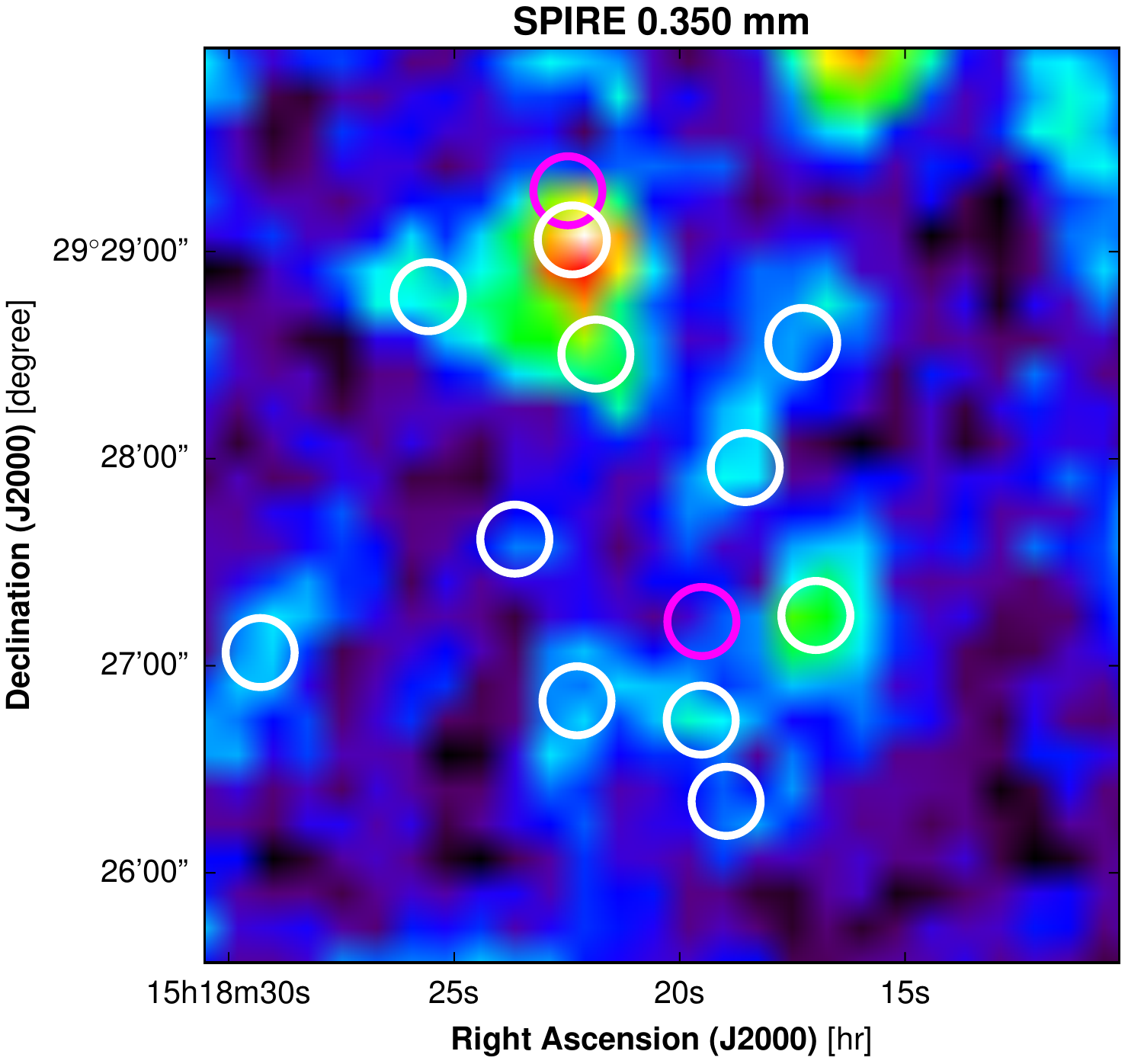}
\hspace{0.4cm}
\includegraphics[height=3.9cm]{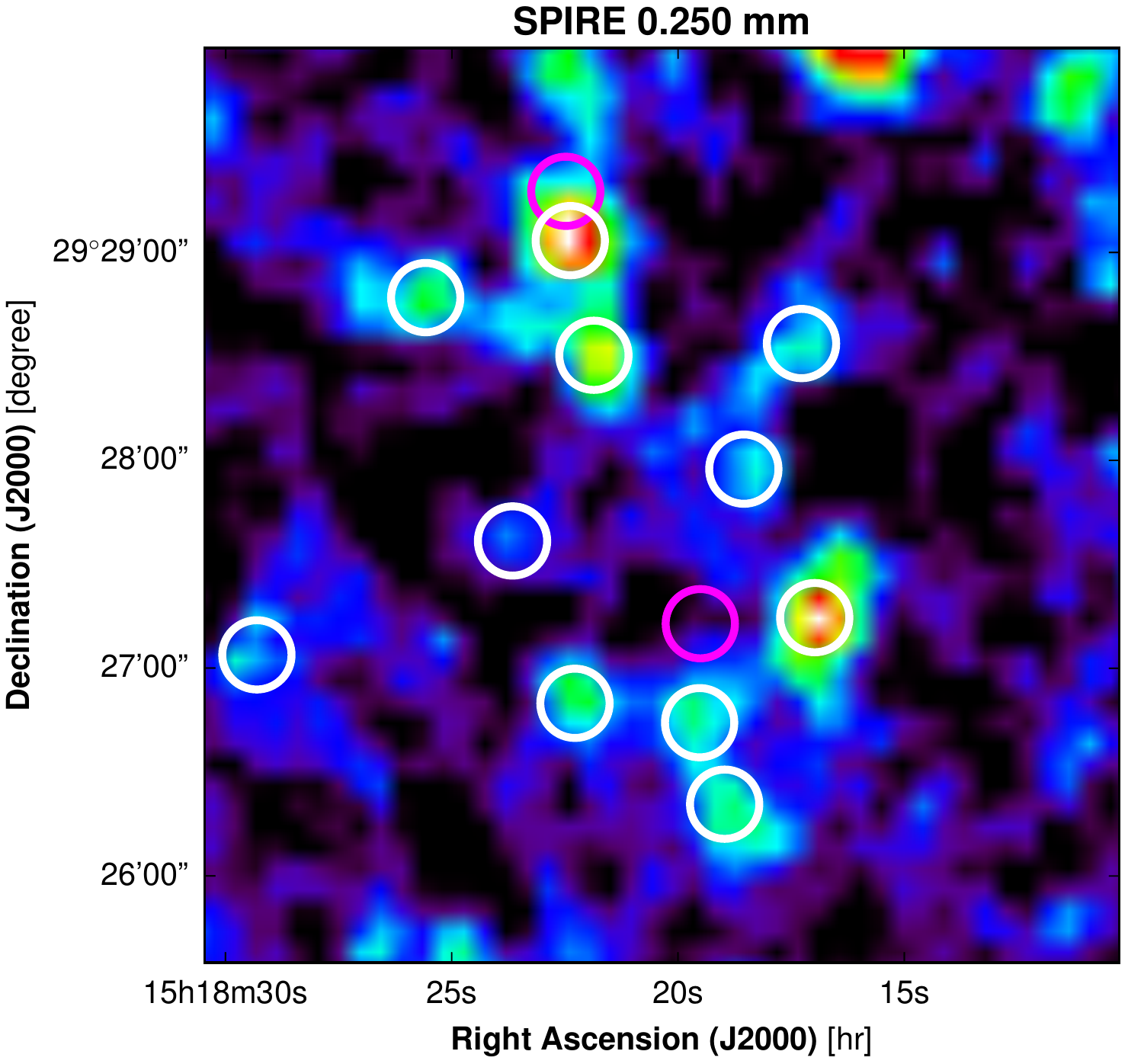}
\caption{{\footnotesize Multiwavelength data set of \mbox{PSZ1\,G045.85+57.71} sky region (the displayed region is 4.4~arcmin wide). The considered instrument is indicated on the top of each map. The maps are smoothed and their range is adapted for visualization purposes. The 10 arcsec radius circles show the point source locations,  in magenta for radio sources (table \ref{tab:Radio_ps}) and in white for submillimeter sources (table \ref{tab:Submm_ps_flux}).}}
\label{fig:point_sources}
\end{figure*}
\begin{table*}[h]
\begin{center}
\begin{tabular}{ccccccccc}
\hline
\hline
Source & $250~\mu$m source position & $250~\mu$m & $350~\mu$m & $500~\mu$m & 1.15 mm & 2.05 mm & 2.05 mm & RMS\\
 &  & [mJy] & [mJy] & [mJy] & [mJy] & measured [mJy] & expected [mJy] & [mJy]\\
\hline
SMG1 & 15:18:22.382, +29:29:03.42 & $43.6\pm 2.4$ & $51.4\pm 3.4$ & $61.2\pm 4.8$ & $1.9\pm 2.2$ & $0.33\pm  0.42$& $0.27\pm  0.09$ & 0.42\\
SMG2 & 15:18:16.978, +29:27:14.60 & $36.7\pm 2.3$ & $27.3\pm 3.0$ & $13.6\pm 3.5$ & $-2.1\pm 1.9$ & **& $0.04\pm  0.03$ & 0.36\\
SMG3 & 15:18:21.859, +29:28:30.33 & $27.0\pm 2.4$ & $-1.9\pm 9.3$ & $2.1\pm 9.8$ & $-1.3\pm 2.1$ & **& $0.01\pm  0.01$ & 0.38\\
SMG4 & 15:18:22.277, +29:26:49.87 & $24.1\pm 2.5$ & $9.5\pm 9.3$ & $3.6\pm 9.1$ & $-0.2\pm 1.9$ & $-0.22\pm  0.36$& $0.07\pm  0.05$ & 0.36\\
SMG5 & 15:18:25.577, +29:28:46.96 & $19.4\pm 2.2$ & $-4.1\pm 9.5$ & $0.6\pm 9.9$ & $-1.8\pm 2.2$ & $0.27\pm  0.43$& $0.01\pm  0.02$ & 0.43\\
SMG6 & 15:18:17.270, +29:28:33.72 & $18.2\pm 2.4$ & $14.6\pm 3.4$ & $-0.6\pm 9.3$ & $2.3\pm 2.1$ & $0.08\pm  0.38$& $0.01\pm  0.01$ & 0.39\\
SMG7 & 15:18:18.969, +29:26:20.75 & $14.0\pm 2.2$ & $-2.6\pm 9.2$ & $9.0\pm 10.2$ & $1.7\pm 2.1$ & $0.26\pm  0.39$& $0.09\pm  0.05$ & 0.39\\
SMG8 & 15:18:23.657, +29:27:36.72 & $13.9\pm 2.2$ & $-1.9\pm 9.4$ & $-6.9\pm 9.8$ & $-0.8\pm 1.9$ & $0.21\pm  0.36$& $0.04\pm  0.05$ & 0.36\\
SMG9 & 15:18:29.306, +29:27:03.83 & $13.9\pm 2.3$ & $17.3\pm 3.1$ & $9.7\pm 9.8$ & $3.2\pm 3.3$ & $-0.29\pm  0.48$& $0.02\pm  0.01$ & 0.48\\
SMG10 & 15:18:19.522, +29:26:44.27 & $11.1\pm 2.1$ & $9.7\pm 3.1$ & $6.7\pm 10.0$ & $1.1\pm 1.9$ &  $0.51\pm  0.38$&  $0.04\pm  0.02$ & 0.36\\
SMG11 & 15:18:18.545, +29:27:57.47 & $8.1\pm 2.2$ & $-4.0\pm 9.0$& $-3.1\pm 9.7$ & $0.7\pm 1.8$ & **& $0.01\pm  0.01$ & 0.36\\
\hline
\hline
\end{tabular}
\end{center}
\caption{{\footnotesize Positions and fluxes of the 11 submillimeter sources identified in the $4.4 \times 4.4$ $\mathrm{arcmin^2}$ field around \mbox{PSZ1\,G045.85+57.71}, measured by fitting Gaussian models to the {\it Herschel} maps at each wavelength as described in Sec. \ref{sec:point_source}. The 260~GHz NIKA map is also used to constrain each source SED at low frequency. Fluxes at 150~GHZ, which are not available because of the tSZ contamination are denoted by double stars **. The expected fluxes at 150~GHz are computed by integrating the estimated SED in the NIKA bandpass. The final column corresponds to the NIKA 150~GHz band RMS noise at the source locations.}}
\label{tab:Submm_ps_flux}
\end{table*}

\subsection{Previous SZ observations of \mbox{PSZ1\,G045.85+57.71}}\label{sec:previous_sz}
%========== Planck and AMI data
\mbox{PSZ1\,G045.85+57.71} has been identified by \planck\ with a S/N of 5.06. It is a member of the early \planck\ SZ catalog (\citealt{Planck_cata}) and its detection has been confirmed in the second catalog release (\citealt{Planck_cata2}).
The \planck\ tSZ map of \mbox{PSZ1\,G045.85+57.71} is shown in the left panel of figure \ref{fig:Planck_XMM}. It has been obtained by extracting a patch of the \planck\ full sky $y$-map using a Gnomonic projection (\citealt{ymap_planck}). The patch used in the ICM analysis is centered on the cluster coordinates and is 1.7 degree wide. Its integrated Compton parameter estimated at $\rm{R_{500}}$ is given in the \planck\ catalog by \mbox{$\rm{Y_{500}} = 8.21^{+1.73}_{-1.70} \times 10^{-4} \, \mathrm{arcmin^2}$} (\citealt{PSZ1_updated})\footnote{$\rm{R_{\Delta}}$ is the radius within which the mean cluster density is equal to $\Delta$ times the critical density of the Universe at the cluster redshift \mbox{$\rho_c = \frac{3H(z)^2}{8\pi G}$}.}. The corresponding cylindrical integrated Compton parameter at $\rm{5R_{500}}$ is given by multiplying this value by 1.79, when assuming the universal pressure profile of \citealt{universal}. This estimation has been compared with the integrated Compton parameter found by aperture photometry on the \planck\ $y$-map, \mbox{$\rm{Y_{5R500}} = 1.28\pm 0.57 \times 10^{-3} \, \mathrm{arcmin^2}$}. The error on the estimated value was computed by performing the same aperture photometry measurement on the \planck\ map randomly around the cluster, where the noise is homogeneous. The estimate of the cluster integrated Compton parameter computed by aperture photometry on the map is therefore compatible with that given in the \planck\ catalog (\citealt{PSZ1_updated}). The \planck\ estimated hydrostatic mass assuming the best-fit Y-M scaling relation of (\citealt{universal}) as a prior was found to be $\rm{M_{500}} = 7.936^{+0.894}_{-0.962} \times 10^{14}M_{\odot}$ (\citealt{PSZ1_updated}). The uncertainties on this cluster mass estimation does not take into account the intrinsic scatter of the scaling relation or systematic errors coming from the data selection for the fit of the scaling relation.  A tSZ follow-up of this cluster has been made at 15 GHz by the Arcminute Microkelvin Imager (AMI) at a slightly better resolution (3 arcmin compared to the \planck\ beams of 5-10 arcmin; see \citealt{AMI_followup}). These AMI observations provide a joint estimation of both the characteristic angular size $\theta_s$ and the integrated Compton parameter $\rm{Y_{tot}}$. The latter is equivalent to the \planck\ estimator $\rm{Y_{5R500}}$ to within 5\% if we assume the universal pressure profile with universal parameter values and the concentration parameter $c_{500} = 1.177$ (\citealt{universal,AMI_followup}). The results derived by AMI are compatible with the \planck\ results. The combination of both \planck\ and AMI constraints gives an integrated Compton parameter estimation at \mbox{$\rm{Y_{5R500}^{\rm{Planck/AMI}}} = 1.47 \pm 0.51 \times 10^{-3} \, \mathrm{arcmin^2}$} thus improving the \planck\ estimation by about 10\%. This estimation is used, along with the NIKA data, to give a first estimate of the radial pressure profile of \mbox{PSZ1\,G045.85+57.71}.

\subsection{Point source contamination}\label{sec:point_source}

\begin{figure}[h]
\centering
\includegraphics[height=6.6cm]{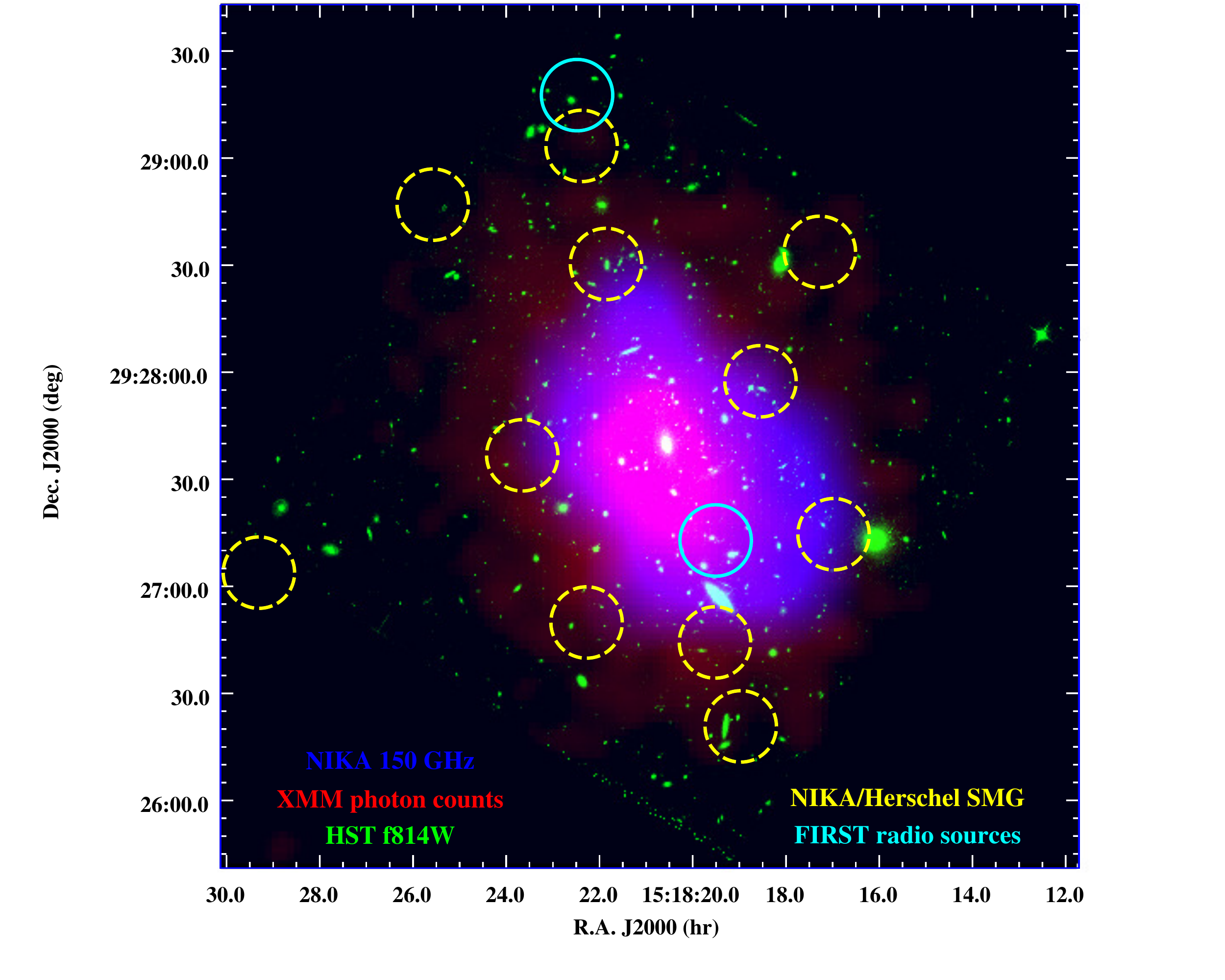}
\caption{{\footnotesize Multiprobe combined map of \mbox{PSZ1\,G045.85+57.71}. \textbf{Blue:} NIKA tSZ surface brightness map giving an estimate of the ICM pressure distribution. \textbf{Red:} XMM-{\it Newton} X-ray photon count map tracing the electronic density squared. \textbf{Green:} Hubble Space Telescope image of the cluster using the F814W filter and showing the cluster galaxy locations. The solid cyan and dashed yellow circles give respectively the radio and submillimeter point sources identified in the field of view.}}
\label{fig:rgb_image}
\end{figure}
%========== Identified point sources (radio and submm)
As has been shown in \citealt{SayersPointSource} and \citealt{MACSJ1424NIKA}, the point source contamination of the tSZ signal has to be studied carefully to avoid significant bias in the ICM characterization. The NRAO VLA Sky Survey (NVSS; \citealt{Condon1998}) and Faint Images of the Radio Sky at Twenty-Centimeters survey (FIRST; \citealt{Becker1995}), which cover the northern sky at 1.4~GHz, has enabled the detection of two radio sources identified as galaxies in the region observed by NIKA. The first source, hereafter RS1, is located in the southwest region of \mbox{PSZ1\,G045.85+57.71} at about 40 arcsec from the X-ray center. The second source, named RS2 in the following, is located at about 2 arcmin toward the north of the X-ray center. The fluxes of RS1 and RS2 are only significant at 1.4 GHz and their values are reported in table \ref{tab:Radio_ps}. We follow the method detailed in \citealt{MACSJ1424NIKA} and model the radio source spectral energy distributions (SED) by a power law, $F_{\nu} = F_{\rm{1~GHz}}\left(\frac{\nu}{\rm{1~GHz}}\right)^{\alpha_{\rm{radio}}}$, to estimate their expected fluxes in the NIKA
bandpasses. As there is no other measurement of RS1 and RS2 fluxes referenced in other catalogs at different frequencies, we only constrain the $F_{\rm{1~GHz}}$ parameter and let the spectral index as a random variable following a Gaussian pdf centered on -0.7 and with a standard deviation of 0.2, which is typical of radio galaxies (see \citealt{spectral_index}). The SEDs are thus simulated by computing Monte Carlo realizations of the radio source fluxes and spectral index within their error bars. The generated SEDs are then integrated within the NIKA bandpasses to predict the expected flux at 150 and 260~GHz. The results are reported in table \ref{tab:Radio_ps_flux}. Given the mean RMS noise at the identified radio source locations we therefore conclude that their contamination at the NIKA frequencies is negligible.\\
\indent We also consider the {\it Herschel} SPIRE (\citealt{SPIRE}) data to identify submillimeter point sources and compute their expected fluxes at 150~GHz. Eleven sources are found in the region observed by NIKA thanks to the SPIRE $250~\mu$m catalog. The corresponding sources in the other channels ($350~\mu$m and $500~\mu$m) are inferred from their respective positions in the $250~\mu$m map. Following the procedure detailed in \citealt{MACSJ1424NIKA}, the fluxes of the sources are measured in the three SPIRE channels by fitting Gaussian functions at the source positions with a fixed FWHM given by the corresponding {\it Herschel} resolution in each channel (35.2, 23.9, 17.6 arcsec at 500, 350, and 250 $\mu$m respectively). A local background is also fitted. Uncertainties on the flux of the sources are inferred by computing the dispersion of fluxes estimated by fitting the same Gaussian functions at random positions, where the noise is homogeneous. The 260~GHz NIKA map was also used to constrain the SED slope at low frequency for each source. The computed submillimeter point source fluxes are presented in table \ref{tab:Submm_ps_flux}. The estimated fluxes corresponding to identified sources are compatible with the values reported in the {\it Herschel} catalog. A modified blackbody spectrum
\begin{equation}
F_{\nu} = A_0\left(\frac{\nu}{\nu_0}\right)^{\beta_{\mathrm{dust}}}B_{\nu}(T_{\mathrm{dust}})
\end{equation}
was used to model the SED of the identified submillimeter point sources from their estimated fluxes. In this model, $A_0$ is a normalization, $\nu_0$ a reference frequency, $\beta_{\mathrm{dust}}$ the dust spectral index, and $T_{\mathrm{dust}}$ the dust temperature. A Markov Chain Monte Carlo (MCMC) analysis was performed to compute the best-fit SED model for each source. The estimated SEDs are then integrated in the NIKA 150~GHz bandpass to quantify the point source contamination at this frequency. The computed fluxes at 150~GHz are reported in table \ref{tab:Submm_ps_flux} and take the SPIRE color correction into account. These results show that the submillimeter point source contamination at 150~GHz is one order of magnitude below the corresponding NIKA RMS noise at this frequency. We therefore conclude that this contamination is negligible and do not consider either radio or submillimeter point sources in the ICM characterization developed in Sec. \ref{sec:Radial_pressure_reconstruction}.

\subsection{XMM-{\it Newton} observations}\label{sec:XMM}
%========== XMM-Newton follow-up
An X-ray follow-up of the \planck-discovered clusters has been undertaken since Spring 2010 (\citealt{XMM_followup}). Thus, \mbox{PSZ1\,G045.85+57.71} has been observed for $\sim 24$ ks by the EPIC instruments during XMM-{\it Newton} revolution 2303 (2012 July 6). The basic data reduction (\emph{i.e.}, production of cleaned and calibrated event files, vignetting correction, point source removal, and the production of associated background data sets) followed the procedures described in \citealt{MACSJ1424NIKA} and references therein. About 15~ks of exposure time remained after the data cleaning.\\
\indent The X-ray image shown in Fig.~\ref{fig:Planck_XMM}, combining the data from all three EPIC detectors, was produced as described in \citealt{boe10}. Here the background subtraction, undertaken for each detector separately, was obtained from a model fit to an image with all sources (including the cluster) excised. The model, consisting of smoothly varying vignetted and unvignetted components, was normalized to the surface brightness in the outer cluster-free regions of the image and was then subtracted to the data.

\subsection{Multiprobe combined map of \mbox{PSZ1\,G045.85+57.71}}\label{sec:map_combi}

A combined map of observations of  \mbox{PSZ1\,G045.85+57.71} with various probes is shown in Fig. \ref{fig:rgb_image} as a conclusion of this section. The identified point source positions are shown by 10 arcsec radii circles in cyan and yellow for the radio and submillimeter sources, respectively. The  Hubble Space Telescope (HST) observed \mbox{PSZ1\,G045.85+57.71} on June 2015 with an exposure time of 1200 s (\citealt{HST_followup}). The galaxy distribution (shown in green on the figure) identified by the HST using the F814W filter follows an elliptical structure that is consistent with the ICM morphology measured by XMM-{\it Newton} and NIKA in red and blue, respectively. Although, the SZ and X-ray peak position are well aligned on the map, we cannot conclude on the cluster relaxation state because the NIKA S/N at \mbox{$R_{500}$} is not high enough. However, the XMM-{\it Newton} observations along with the galaxy distribution reveal that this cluster has a significant elliptical morphology with a projected major axis oriented from the southwest of the X-ray center to the northeast. The next section describes how the \mbox{PSZ1\,G045.85+57.71} ICM is modeled in the following multiprobe analysis given the morphology constraints that we can derive from the NIKA tSZ surface brightness map.

%#####################################################################################
%##########################       Justification of the spherical model assumption        #################%#####################################################################################

\section{Modelization of the ICM}\label{sec:ellipticity}

\subsection{Parametric modeling}\label{sec:modeling}
%---------- What we do
The  combination of NIKA and X-ray data, such as the XMM-{\it Newton} data, can bring new insights into the ICM thermodynamics reconstruction. Indeed, the electronic density within the ICM is low enough to consider it as an ideal gas for which the temperature is simply given by the ratio between the NIKA estimated pressure and the XMM-{\it Newton} estimated density at each point of the ICM. This method allows us to constrain the temperature profile of a galaxy cluster with almost no spectroscopic information. Indeed, the mean ICM temperature that is needed to deproject the cluster density profile can be estimated with only few spectroscopic data, whereas a temperature profile deprojection from spectroscopy measurements is time consuming at high redshift.\\
In the context of spherical cluster symmetry, we can model the ICM by the standard pressure and density models used in previous studies.
%---------- Model
The radial distribution of the cluster electronic pressure is modeled by a generalized Navarro-Frenk-White (gNFW) profile (\citealt{gnfw_prof}), given by
\begin{equation}
        P_e(r) = \frac{P_0}{\left(\frac{r}{r_p}\right)^c \left(1+\left(\frac{r}{r_p}\right)^a\right)^{\frac{b-c}{a}}},
\label{eq:gNFW}
\end{equation}
where $P_0$ is a normalization constant, $r_p$ is a characteristic radius, and $a$, $b,$ and $c$ give the slope of the profile at intermediate, large, and small radii, respectively. The electronic density was modeled by a simplified Vikhlinin model (SVM) (\citealt{SVM_prof}) given by
\begin{equation}
        n_e(r) = n_{e0} \left[1+\left(\frac{r}{r_c}\right)^2 \right]^{-3 \beta /2} \left[ 1+\left(\frac{r}{r_s}\right)^{\gamma} \right]^{-\epsilon/2 \gamma},
\label{eq:SVM}
\end{equation}
where $n_{e0}$ is the central density, $r_c$ is the core radius, and $r_s$ the transition radius at which an additional steepening in the profile occurs. The $\beta$ parameter gives the inner profile slope and $\epsilon$ the outer profile slope. The $\gamma$ parameter describes the width of the transition in the profile. In the following, we fix the $\gamma$ value at 3 since it provides a good fit to all clusters considered in the analysis of (\citealt{gamma_val}).\\
\indent Models for both temperature and entropy profiles are naturally deduced from pressure and density if we consider the ICM as an ideal gas, $k_B T_e(r) = \frac{P_e(r)}{n_e(r)}$ and $K(r) =  \frac{P_e(r)}{n_e(r)^{5/3}}$, where $k_B$ is the Boltzmann constant. Assuming hydrostatic equilibrium, the total mass enclosed within the radius $r$ is then given by
\begin{equation}
M_{\rm HSE}(r) = -\frac{r^2}{\mu_{gas} m_p n_e(r) G} \frac{dP_e(r)}{dr},
\label{eq:hse_mass}
\end{equation}
where $\mu_{\rm{gas}} = 0.61$ is the mean molecular weight of the gas, $m_p$ the proton mass, and $G$ the Newton constant.
\begin{figure}[h]
\centering
\includegraphics[height=5.7cm]{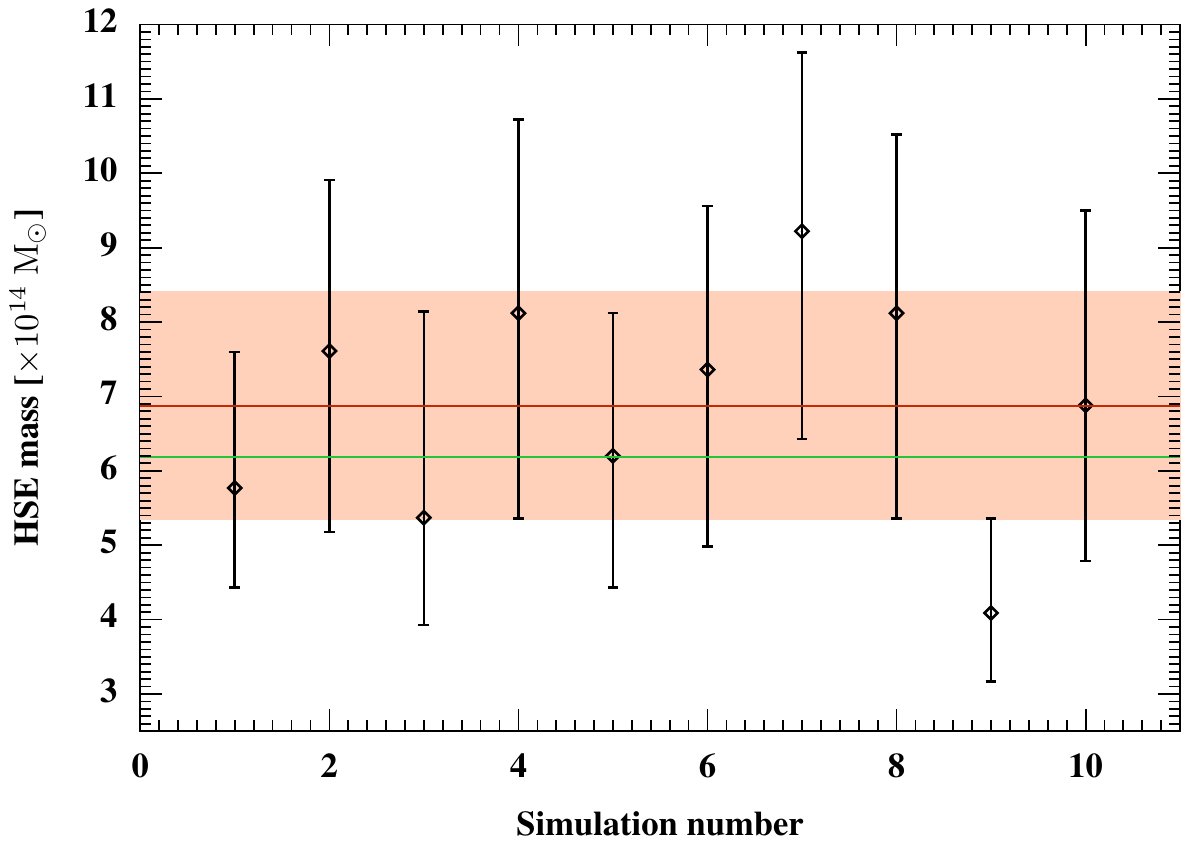}
\caption{{\footnotesize Mass estimates (black diamonds) for the simulated triaxial clusters with their major axis orthogonal to the line of sight. The red line indicates the mean of the recovered distribution and the salmon region is its dispersion. The input mass given by the gNFW and SVM models considered for the simulation is represented by the green line.}}
\label{fig:ellipticity_mass}
\end{figure}

\subsection{Cluster mass estimation from a MCMC analysis}\label{sec:MCMC_old}
%---------- Mass estimation from a MCMC analysis
In order to estimate the mass of the cluster, one must first evaluate both the density and pressure profiles describing the ICM. This ICM characterization method has been presented in detail in (\citealt{CLJ1227NIKA}) and we only explain the key points of the analysis in this section. We use the information contained in the 150 GHz tSZ surface brightness map to constrain the gNFW profile parameters  from a MCMC sampling of the full parameter space. The main advantages of this approach are that we can use all the information contained in the map to constrain the parameters of interest, marginalize over nuisance parameters, and identify parameter correlations during the MCMC sampling. The inner slope of the pressure profile cannot be constrained at the considered cluster redshift because of the NIKA beam dilution. Therefore, all the gNFW parameters are kept free except for $c$, which is 
fixed to the value estimated by \citealt{Planck_pressure_prof} at $c=0.31$. At each step of the procedure, a set of parameters is generated and defines the corresponding pressure radial profile. The latter is integrated along the line of sight to compute a Compton parameter angular profile from which we derive a tSZ surface brightness map model $M_{\mathrm{model}}$ at 150~GHz and an integrated Compton parameter $Y_{\rm{tot}}^{\rm{model}}$ evaluated up to $5R_{500}$. The relativistic corrections in the tSZ spectrum (eq. \ref{eq:tSZ_spectrum}) are computed from the temperature estimate given by the ratio between the cluster pressure profile and its density profile. Both $M_{\mathrm{model}}$ and $Y_{\rm{tot}}^{\rm{model}}$ are then compared to the observed 150~GHz tSZ surface brightness map ($M_{\mathrm{data}}$) and integrated Compton parameter ($Y_{\rm{tot}}^{\rm{data}}$) using the following Gaussian likelihood model:
\begin{equation}
\begin{tabular}{rl}
        $-2 \mathrm{ln} \, \mathscr{L} $  & $ =\chi^2_{\mathrm{SZ~map}} + \chi^2_{\mathrm{Y_{tot}}}$\\[0.2cm]
         &$ =\sum_{i=1}^{N_{\mathrm{pixels}}} [(M_{\mathrm{data}} - M_{\mathrm{model}})^T C_{\mathrm{NIKA}}^{-1} (M_{\mathrm{data}} - M_{\mathrm{model}})]_i $ \\[0.2cm]
         & $+ \left( \frac{Y_{\rm{tot}}^{\rm{data}} - Y_{\rm{tot}}^{\rm{model}}}{\sigma_{\mathrm{data}}} \right)^2$
\label{eq:likelihood}
\end{tabular}
.\end{equation}
The MCMC sampling procedure also marginalizes over nuisance parameters such as the zero level of the NIKA map and the calibration coefficient uncertainty. The sampling stops when the convergence criteria given by \citealt{convergence_MCMC} is reached for all the fitted parameters. The final likelihood function marginalized distributions are eventually given by the remaining chain points after the burn-in cutoff, which discards the first 10\% of each chain. These distributions are then used to compute the gNFW parameter constraints that define the best ICM pressure profile. Both density and pressure profiles are then used to compute a mass profile using equation \ref{eq:hse_mass} from which we can derive the cluster total mass $M_{\mathrm{500}}$.

\subsection{Impact of the departure from sphericity on the ICM thermodynamic reconstruction}\label{sec:Impact_mass}

%========== Ellipticity
A significant amount of disturbed clusters that are characterized, for instance, by the presence of substructures in the ICM, unvirialized ICM, or merging events, are identified at high redshift by high angular resolution observations.\\
\indent In this context, describing the ICM by a spherical model may add dispersion and bias on ICM thermodynamic constrains and galaxy cluster mass estimations. In particular, this is the case if the intrinsic deviation from sphericity is significant, given the residual noise properties measured on the map.\\
\indent This section describes the morphology analysis made on both XMM-{\it Newton} X-ray photon count map and NIKA tSZ surface brightness map at 150 GHz to check the possibility to recover the cluster ellipticity in the individual maps. We then describe the analysis made on simulated tSZ surface brightness maps at 150 GHz to study whether a spherical model is appropriate to derive \mbox{PSZ1\,G045.85+57.71} ICM thermodynamic properties from the NIKA and XMM-{\it Newton} observations.

%========== Ellipticity of PSZ1G045
\subsubsection{\mbox{PSZ1\,G045.85+57.71} ellipticity}\label{sec:PSZ1G045_ellipticity}
%---------- with XMM
As shown in figure \ref{fig:Planck_XMM} (right panel), the XMM-{\it Newton} observations of \mbox{PSZ1\,G045.85+57.71} reveals a significant elliptical morphology of the ICM with a projected major axis oriented from the southwest of the X-ray center to the northeast. As the information along the line of sight is lost, we only constrain the length scales of this cluster in the plane of the sky. The ellipticity, defined by $\epsilon = 1 - \frac{b}{a}$, where $a$ and $b$ are the major and minor axes, respectively,  of the considered ellipse, and the orientation of the major axis is estimated by fitting ellipses on iso-number count contours of the XMM-{\it Newton} photon count map. Their respective ellipticity and orientation were computed and show that \mbox{PSZ1\,G045.85+57.71} has a mean ellipticity of $\epsilon_{\rm{XMM}} = 0.33 \pm 0.01$ and a major axis oriented with an angle $\theta_{\rm{XMM}}^{\rm{maj}} = (70 \pm 2)^{\circ}$ with respect to the R.A. axis in the clockwise direction. The uncertainties on both estimations are statistical only.\\
%---------- with NIKA
\indent As the NIKA RMS noise is fairly constant in the cluster region, the same analysis can been carried out on the NIKA tSZ surface brightness map using constant S/N contours from 3.5 to 6.5 with 0.5 steps to fit the ellipses. This analysis shows a much larger dispersion on the estimated ellipticity and major-axis orientation with $\epsilon_{\rm{NIKA}} = 0.4 \pm 0.1$ and $\theta_{\rm{NIKA}}^{\rm{maj}} = (44 \pm 8)^{\circ}$. The given error bars are statistical only and do not take the correlated noise on the map into account. Indeed, the residual correlated noise on the NIKA map can induce noise structures with angular scales larger than the NIKA beam at 150~GHz that may distort the intrinsic ICM projected morphology. It is therefore important to characterize the bias induced by the spherical cluster assumption on the ICM thermodynamic reconstruction. Such a hypothesis will be adapted if the induced bias is negligible with respect to the uncertainty on the estimated constraints due to the residual noise on the NIKA map.

\subsubsection{Compatibility between NIKA SZ observations and the spherical cluster assumption}\label{sec:impact_mass}
\begin{figure*}[h]
\centering
\includegraphics[height=6.2cm]{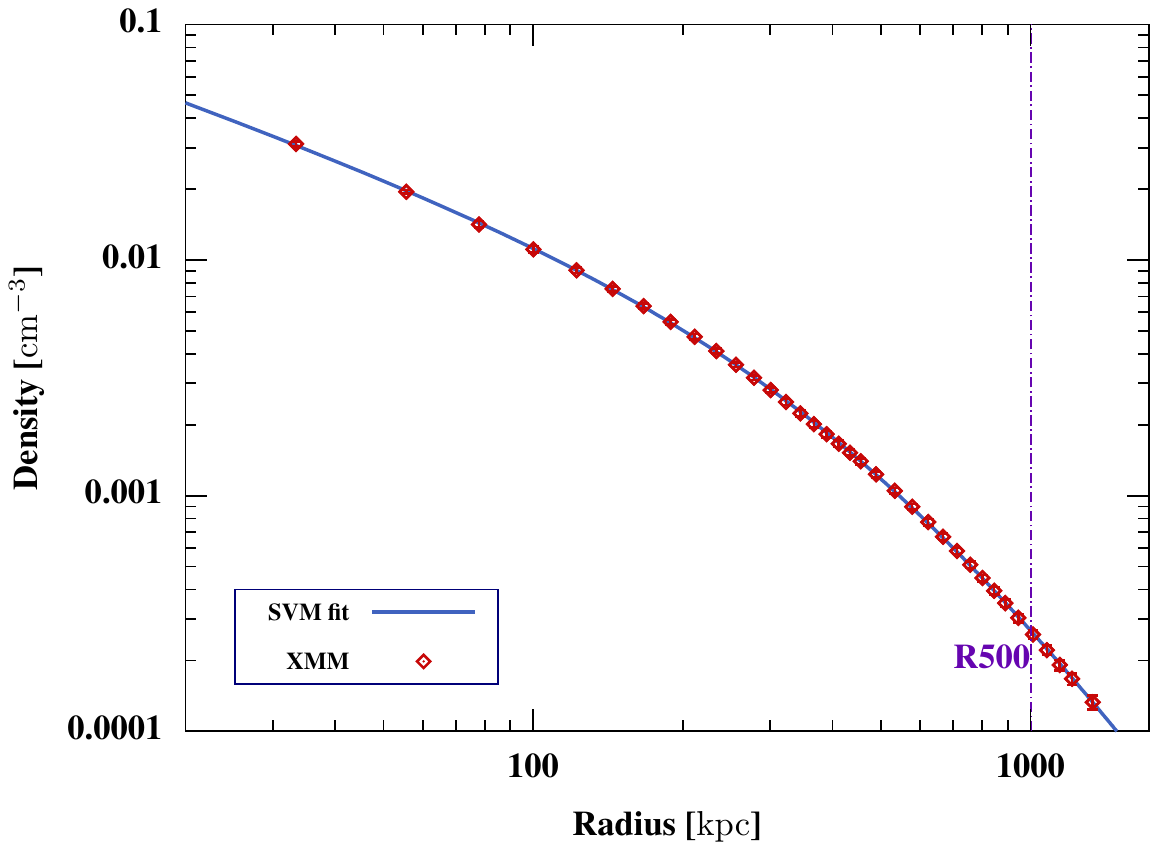}
\hspace{0.7cm}
\includegraphics[height=6.2cm]{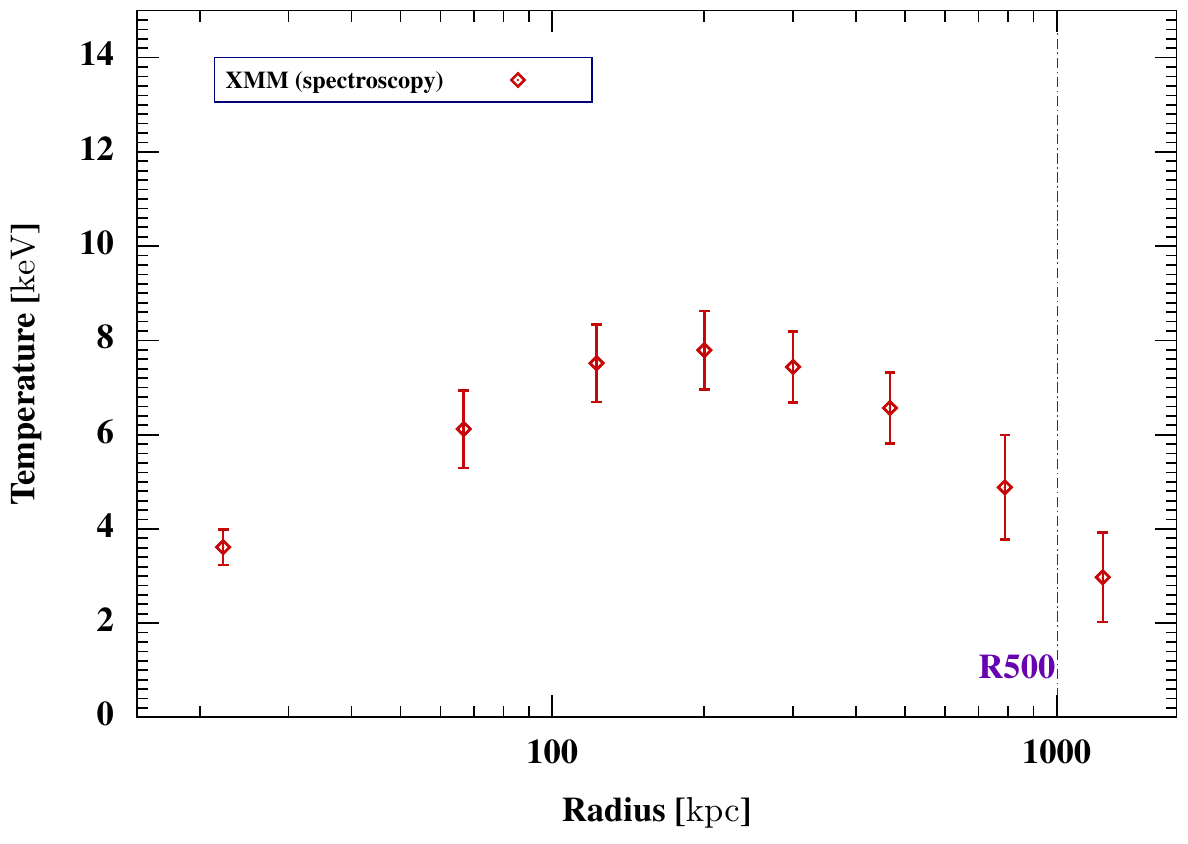}
\caption{{\footnotesize \textbf{Left:} Deprojected electronic density profile derived from the XMM-{\it Newton} data (red dots). The best-fit SVM model (\citealt{SVM_prof}) is given by the blue line. \textbf{Right:} Temperature profile derived from the XMM-{\it Newton} spectroscopy data. The characteristic radius measured from XMM-{\it Newton} data, $R_{\mathrm{500}} = 1013 \pm 13~\mathrm{kpc}$, is represented as a vertical dashed magenta line in both figures.}}
\label{fig:XMM_density}
\end{figure*}
%---------- Impact on the mass reconstruction
Although the ellipticity observed on the NIKA map may be due to residual noise, \mbox{PSZ1\,G045.85+57.71} seems to be intrinsically elliptical as shown by the XMM-{\it Newton} map analysis (see Sec. \ref{sec:PSZ1G045_ellipticity}). Therefore, to see if the spherical model assumption induces a significant bias on the final ICM characterization we choose the cluster total mass estimation as a quantitative indicator of the induced bias.\\
\indent Knowing the cluster projected ellipticity and major-axis orientation from the XMM-{\it Newton} observations, we derive a tSZ surface brightness map from a simulated ellipsoidal cluster presenting similar projected morphological properties. The adopted model is a gNFW pressure profile with a modified radius defined as
\begin{equation}
r = \sqrt{\left(\frac{x \, \mathrm{cos}(\theta) - y \, \mathrm{sin}(\theta)}{a}\right)^2 + \left(\frac{x \, \mathrm{sin}(\theta) + y \, \mathrm{cos}(\theta)}{b}\right)^2 + \left(\frac{z}{c}\right)^2},
\label{eq:modif_rad}
\end{equation}
where $(a, b, c)$ define the axis ratios of the triaxial ICM and $\theta$ is the angle between the major axis and the line of sight. If the $\theta$ angle is different from $90^{\circ}$, the major-axis length has to be increased by a factor $1/\mathrm{sin(\theta)}$ to keep the projected ellipticity unchanged. In the following, we consider the ideal case where the projected cluster ellipticity is equal to its intrinsic one (\emph{i.e.,} for a $\theta$ angle of $90^{\circ}$). 
The integration of this pressure model along the line of sight gives us a simulated tSZ surface brightness map model of an elliptical cluster. The tSZ surface brightness simulated maps are computed by adding residual noise to the modeled tSZ signal using the noise power spectrum derived from the NIKA null maps. The gNFW model parameters are adjusted so that the tSZ peak significance matches the one that we observe on the NIKA 150 GHz map of \mbox{PSZ1\,G045.85+57.71}.\\
\indent We take the best-fit SVM model parameters of the \mbox{PSZ1\,G045.85+57.71} XMM-{\it Newton} data to model the simulated cluster density distribution (see Sec. \ref{sec:X_ray_profiles}) and use Eq. \ref{eq:modif_rad} for the modified radius definition to get an elliptical density distribution.The total mass of the simulated ellipsoidal cluster is estimated with the input pressure and density models in Eq. \ref{eq:hse_mass}.\\
\indent The simulated maps are then used to estimate the cluster total mass using a spherical model as described in Sec. \ref{sec:MCMC_old}. The estimated total masses of all Monte Carlo realizations are reported in figure \ref{fig:ellipticity_mass}. The input mass is shown as a green line while the salmon region indicates the $1\sigma$ dispersion of the reconstructed mass distribution.\\
\indent The reconstructed masses are consistent with the input model mass. We therefore conclude that the mass estimation given by a spherical model is not significantly biased by the cluster ellipticity if the cluster major axis is orthogonal to the line of sight for this residual noise level. The bias induced by the spherical model assumption on the reconstructed ICM thermodynamic properties is therefore negligible compared to the dispersion caused by the residual noise on the NIKA tSZ surface brightness map at 150~GHz.\\
\indent We note however that if the cluster projected ellipticity is significantly different from its intrinsic one, the input gNFW parameter value has to be changed to get a tSZ surface brightness angular profile similar to the observed one. There is in particular a degeneracy between the $r_p$ parameter value, the intrinsic cluster ellipticity, and the orientation of the major axis with respect to the line of sight. The reconstructed masses could therefore tend to be significantly biased because of projection effects (see for example \citealt{Projection_effect}). We choose to develop this discussion in a forthcoming paper dedicated to simulation.

%#####################################################################################
%##########################              MULTI PROBE COMPARISON             ##########################%#####################################################################################

\section{Radial thermodynamical reconstruction}\label{sec:Radial_pressure_reconstruction}

\begin{figure*}[t]
\centering
\includegraphics[height=13.4cm]{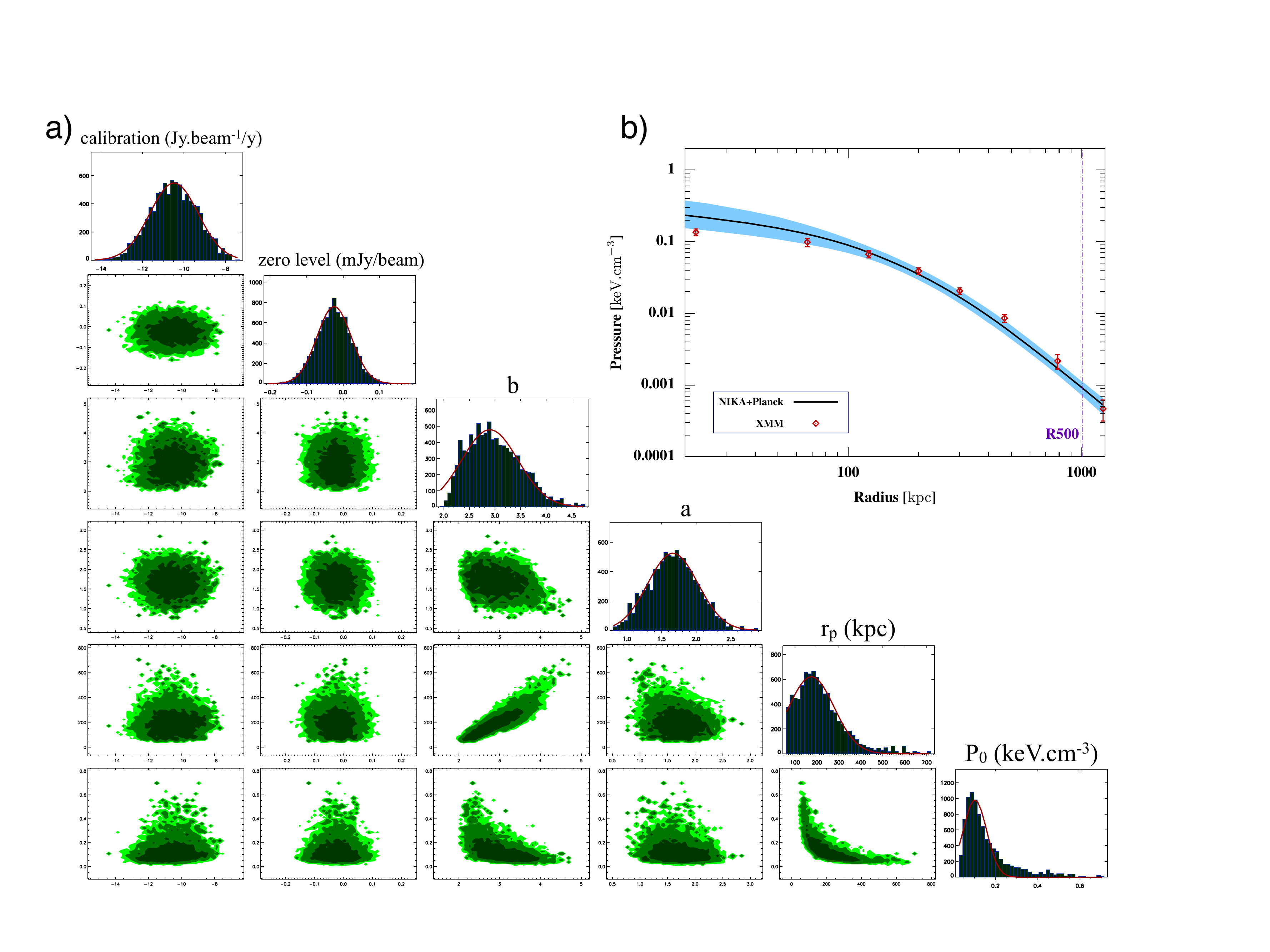}
\caption{{\footnotesize \textbf{Panel a):} Marginalized distributions (diagonal) and 2D correlations (off-diagonal) plots of the parameters of the assumed gNFW model. The MCMC procedure constrains the parameters of interest (from the bottom of the diagonal to the top): $P_0$, $r_p$, $a$, and $b$ and marginalize over the map zero level and $y$-mJy/beam calibration coefficient. \textbf{Panel b):} Maximum likelihood NIKA+\planck/AMI pressure profile (black),  the 1-sigma dispersion (light
blue), and XMM-{\it Newton} constrained pressure profile (red).}}
\label{fig:MCMC1_likeli2D}
\end{figure*}

%--------------Multi-probe analysis
Based on the results presented in the previous section, we assume the spherical symmetry of \mbox{PSZ1\,G045.85+57.71} in the following ICM characterization.\\
This section presents the different methods that have been used to deproject the radial pressure profile of \mbox{PSZ1\,G045.85+57.71}. We first describe how the XMM-{\it Newton} data can be used to recover the electronic density and the gas temperature profiles of \mbox{PSZ1\,G045.85+57.71}. This approach strongly depends on the spectroscopic temperature reconstruction, which is challenging at high redshift because of the cosmological dimming of the X-ray flux. Thus, spectroscopic temperature reconstruction at high redshift requires large integration time to recover the X-ray photon energy spectrum.\\
\indent We then use the procedure described in Sec. \ref{sec:MCMC_old} to estimate the best-fit gNFW pressure profile from the NIKA tSZ surface brightness map at 150~GHz and the \planck/AMI combination of the integrated Compton parameter. This approach gives an estimation of the cluster radial pressure distribution without using spectroscopic information. However, it relies on a specific parametric model, which limits the use of the estimated pressure profile for future studies based on different ICM models.\\
\indent We therefore choose to extract the pressure profile of \mbox{PSZ1\,G045.85+57.71} using a non-parametric spherical model to deproject the NIKA data in the MCMC analysis. Furthermore, instead of considering the \planck\ integrated Compton parameter to constrain the outer slope of the profile we use simultaneously the NIKA tSZ surface brightness map at 150~GHz and the \planck\ MILCA map of the cluster Compton parameter in the MCMC. The whole ICM thermodynamics is then derived by combining the constrained pressure profile and the XMM-{\it Newton} density profile depending only weakly on X-ray spectroscopy.

\subsection{X-ray radial thermodynamic profiles}\label{sec:X_ray_profiles}

\begin{figure*}[h]
\center
\includegraphics[height=7.8cm]{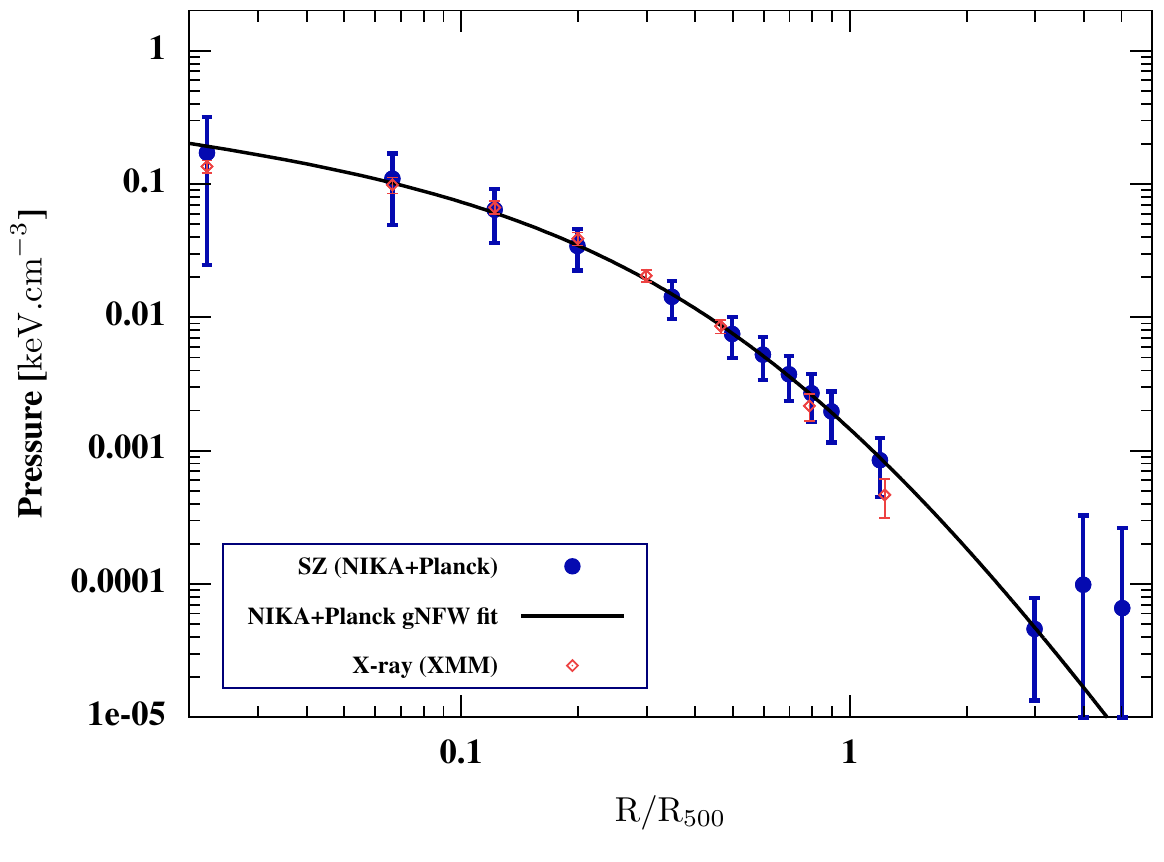}
\caption{{\footnotesize Non-parametric pressure profile (blue) deprojected from the NIKA tSZ surface brightness map and the \planck\ Compton parameter map. The 3 radial bins at 3, 4, and $5R_{500}$ are mostly constrained by the \planck\ data. A gNFW pressure profile model has been fitted on the NIKA+\planck\ deprojected pressure points (black solid line). XMM-{\it Newton} estimated pressure profile (red) based on the deprojected density profile and the temperature estimation from spectroscopic observations. The NIKA/\planck\ and XMM-{\it Newton} estimates are compatible within error bars.}}
\label{fig:Pressure_XMM_NIKA_Planck}
\end{figure*}

%---------- X-ray density profile
Gas density profiles, $n_{\rm e}\,(r)$, produced from the [0.3-2]~keV surface brightness profiles extracted directly from the event files centered on the emission peak, are obtained using the regularized deprojection and PSF-correction procedure described in \citealt{cro06}. Given the strongly constrained electronic density profile from the XMM-{\it Newton} data (shown in red on the left panel of Fig. \ref{fig:XMM_density}), we choose to fit a SVM model on this profile independently and use it in the following multiprobe analysis. The SVM parameters are constrained by minimizing the $\chi^2$ on the deprojected density profile. The best-fit density profile is shown in blue in the figure and perfectly describes the XMM-{\it Newton} measurements. The reduced $\chi^2$ value is estimated at  1.4.

The projected temperature profile was extracted in logarithmically spaced annuli as detailed in \citealt{entropy_REXCESS}. The deprojected radial temperature profile, $T\,(r)$ (shown in red on the right panel of Fig. \ref{fig:XMM_density}), was then obtained by convolving a parameteric model with a response matrix that simultaneously takes into account projection and PSF redistribution, projecting this model, and then fitting it to the projected annular profile. The projection procedure took the bias introduced by fitting isothermal models to multitemperature plasma  into account  (\citealt{maz04, gamma_val}). The computed temperature profile shows a typical cool-core shape, with a central temperature lower than $\sim$4~keV and a peak of $\sim$8~keV at about 200 kpc away from the center.

The gas entropy $K\,(r)$ and pressure $P\,(r)$ distributions were obtained directly from the 3D density and temperature profiles described above. The X-ray mass profile, $M_{\rm HSE}\, (r)$,  derived assuming a spherical gas distribution in hydrostatic equilibrium, was calculated using the Monte Carlo procedure described in \citealt{dem10} and references therein. The high quality of XMM-{\it Newton} spectroscopic data enables us to derive tight constraints on the whole ICM thermodynamics of \mbox{PSZ1\,G045.85+57.71} especially in the cluster core regions. However, such observations are becoming particularly challenging as cluster observations goes toward high redshift.

\subsection{MCMC analysis of the 150 GHz NIKA map with a parametric model}\label{sec:MCMC_old_NIKA}

The entire thermodynamic properties of \mbox{PSZ1\,G045.85+57.71} can also be derived without spectroscopic information by using the cluster pressure profile constrained with the MCMC method described in Sec. \ref{sec:MCMC_old} and the XMM-{\it Newton} deprojected density profile  jointly. The 150~GHz NIKA map is used to constrain both the inner and intermediate parts of the gNFW pressure profile. The \planck/AMI combined estimation of the cluster integrated Compton parameter \mbox{$Y_{5R500}^{\rm{Planck/AMI}} = (1.47 \pm  0.51) \times 10^{-3}~\mathrm{arcmin^2}$} is used in the MCMC analysis to constrain the outer slope of the gNFW pressure profile.\\
\indent The panel a) of Fig. \ref{fig:MCMC1_likeli2D} shows the marginalized distributions (diagonal) and 2D correlations (off-diagonal) of the considered parameters. These distributions are used to compute the gNFW parameter constraints that define the best ICM pressure profile. As shown in Fig. \ref{fig:MCMC1_likeli2D}, the degeneracy between the $r_p$ and $b$ parameters is very strong. All the parameter degeneracies are taken into account when estimating the final pressure profile error bars. The panel b) of Fig. \ref{fig:MCMC1_likeli2D} shows that the NIKA+\planck/AMI pressure profile estimated with this method is compatible with the XMM-{\it Newton} constrained points within error bars.\\
%---------- Where does the constraints come from
\indent The constraints on the pressure profile come almost exclusively from the NIKA and \planck/AMI data. However, we account for relativistic corrections in the tSZ spectrum using the radial temperature profile estimated by combining the deprojected pressure profile with the XMM-{\it Newton} constrained density profile. The overall effect on the final pressure profile is very small compared to the uncertainties coming from residual noise on the NIKA map. The \planck/AMI estimation of the integrated Compton parameter allows the MCMC procedure to avoid models that diverge at large scales, where NIKA is not sensitive and partially breaks the strong degeneracy between the zero level and the $r_p$ parameter (see Fig. \ref{fig:MCMC1_likeli2D}). \planck\ and NIKA are therefore highly complementary to constrain the pressure profile from small to large scales.\\
\indent As shown in the panel b) of Fig. \ref{fig:MCMC1_likeli2D}, the NIKA estimated uncertainties increase in both the cluster core and its outskirts because of the analysis filtering, beam dilution, and the larger RMS noise outside the NIKA FOV, respectively. The most constrained region of the cluster using this method lies therefore between projected angular scales from the X-ray center of about 0.5 and 2 arcmin, which correspond for this cluster redshift to distances from the X-ray center of 200 and 800~kpc, respectively.\\
\indent 

%========== Pressure profile estimation without modeling
\subsection{MCMC analysis based on a non-parametric model}\label{sec:nomodeling}

\begin{figure*}[h]
\centering
\includegraphics[height=4.4cm]{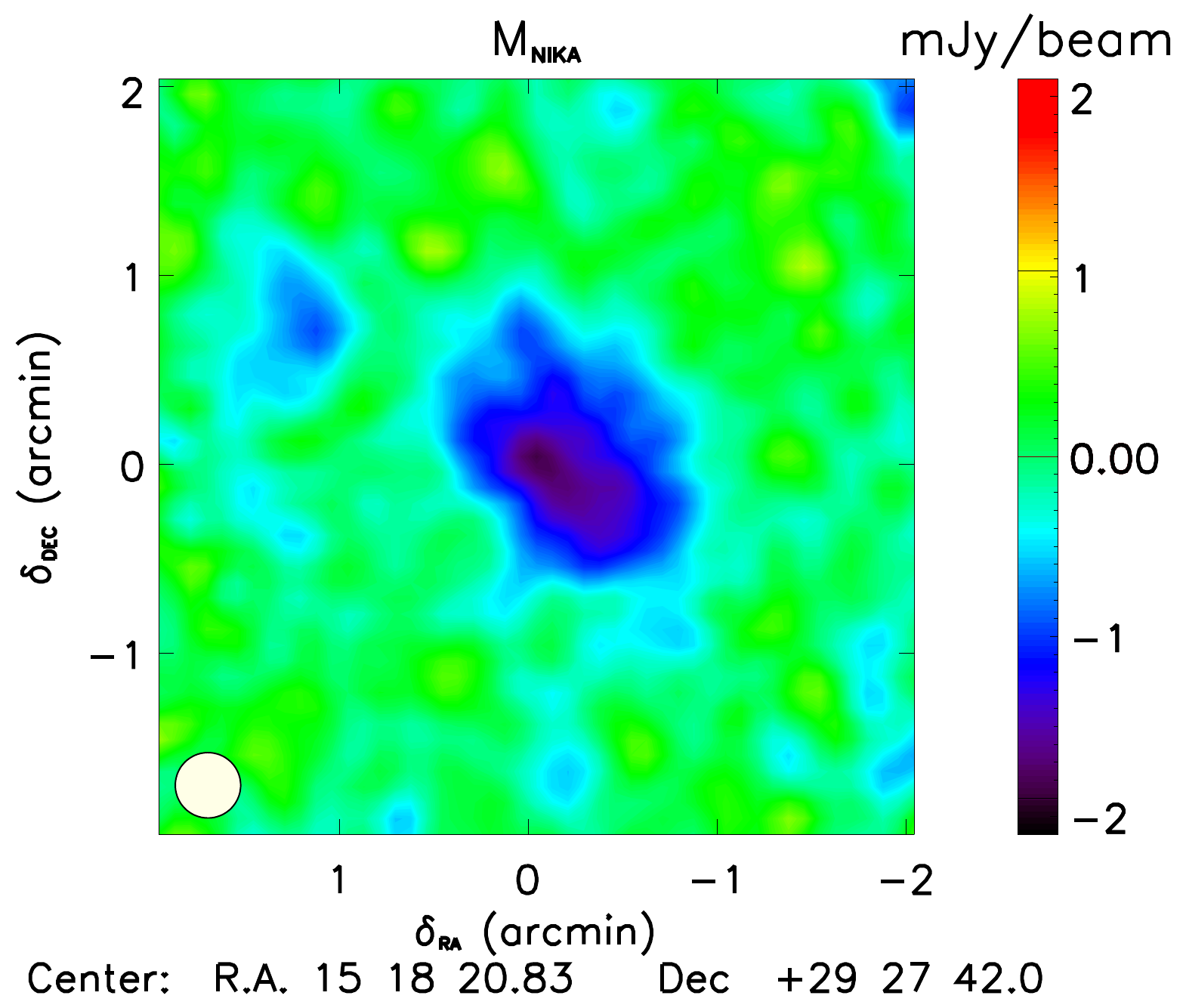}
\hspace{0.7cm}
\includegraphics[height=4.4cm]{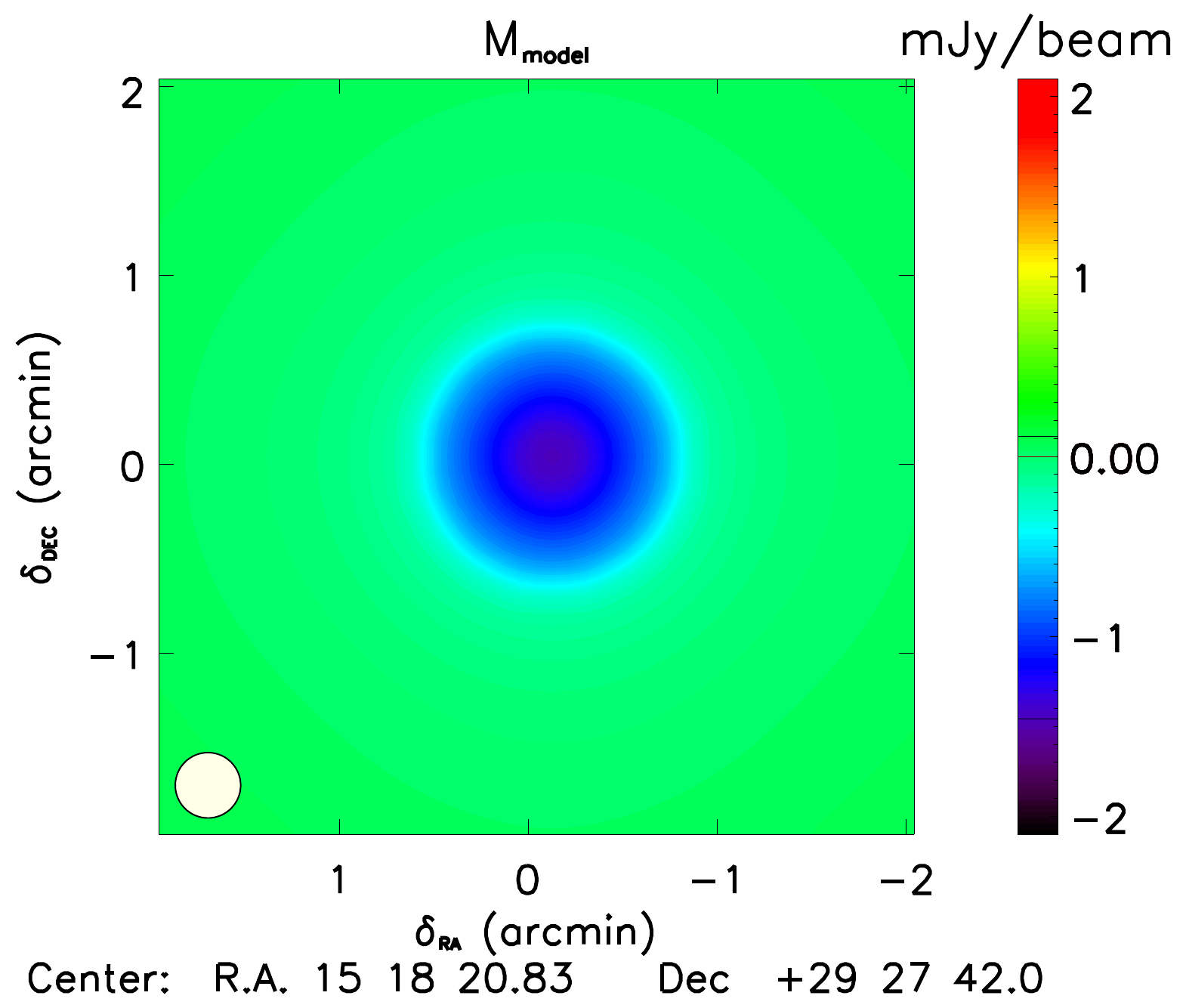}
\hspace{0.8cm}
\includegraphics[height=4.4cm]{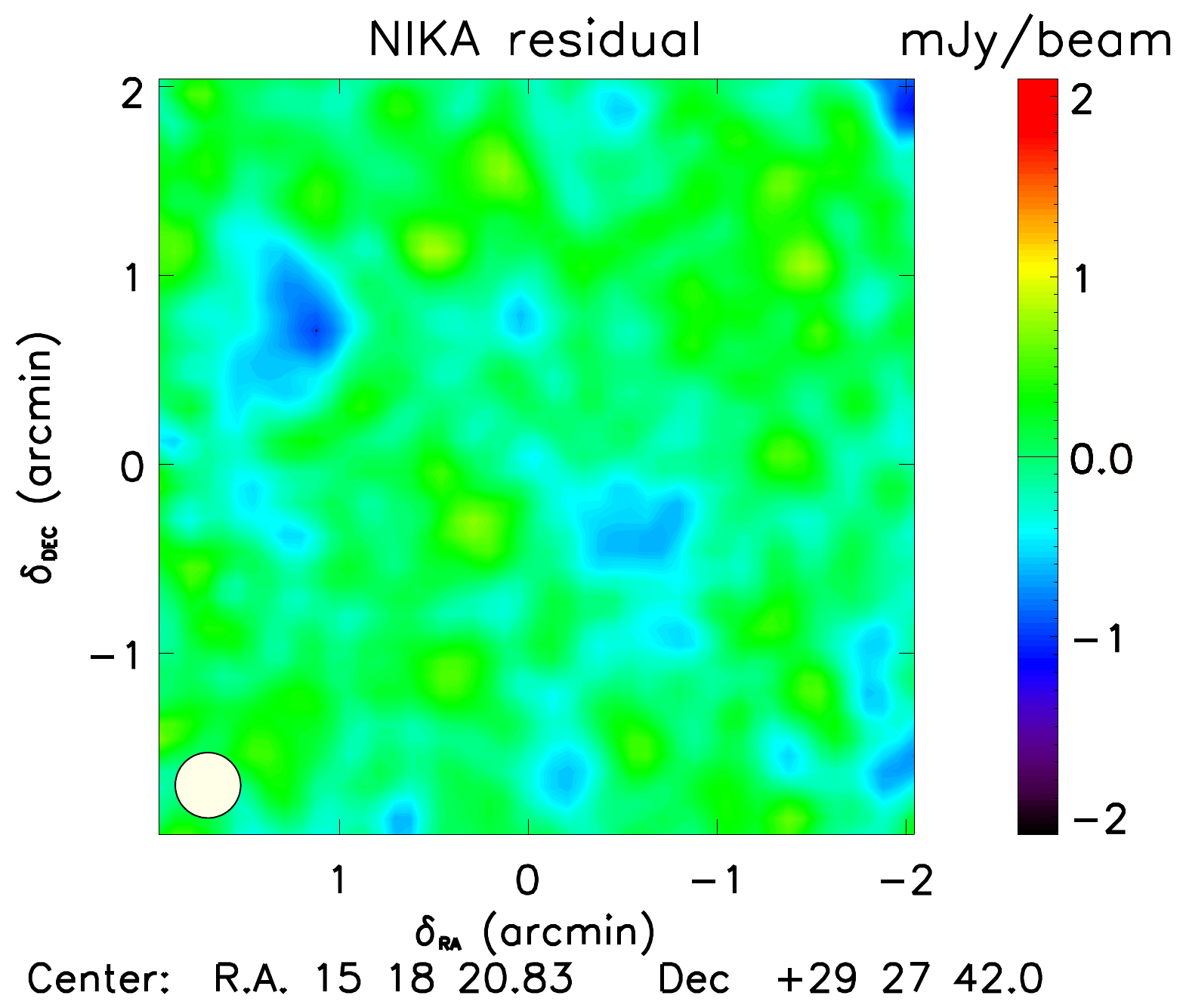}
\includegraphics[height=4.4cm]{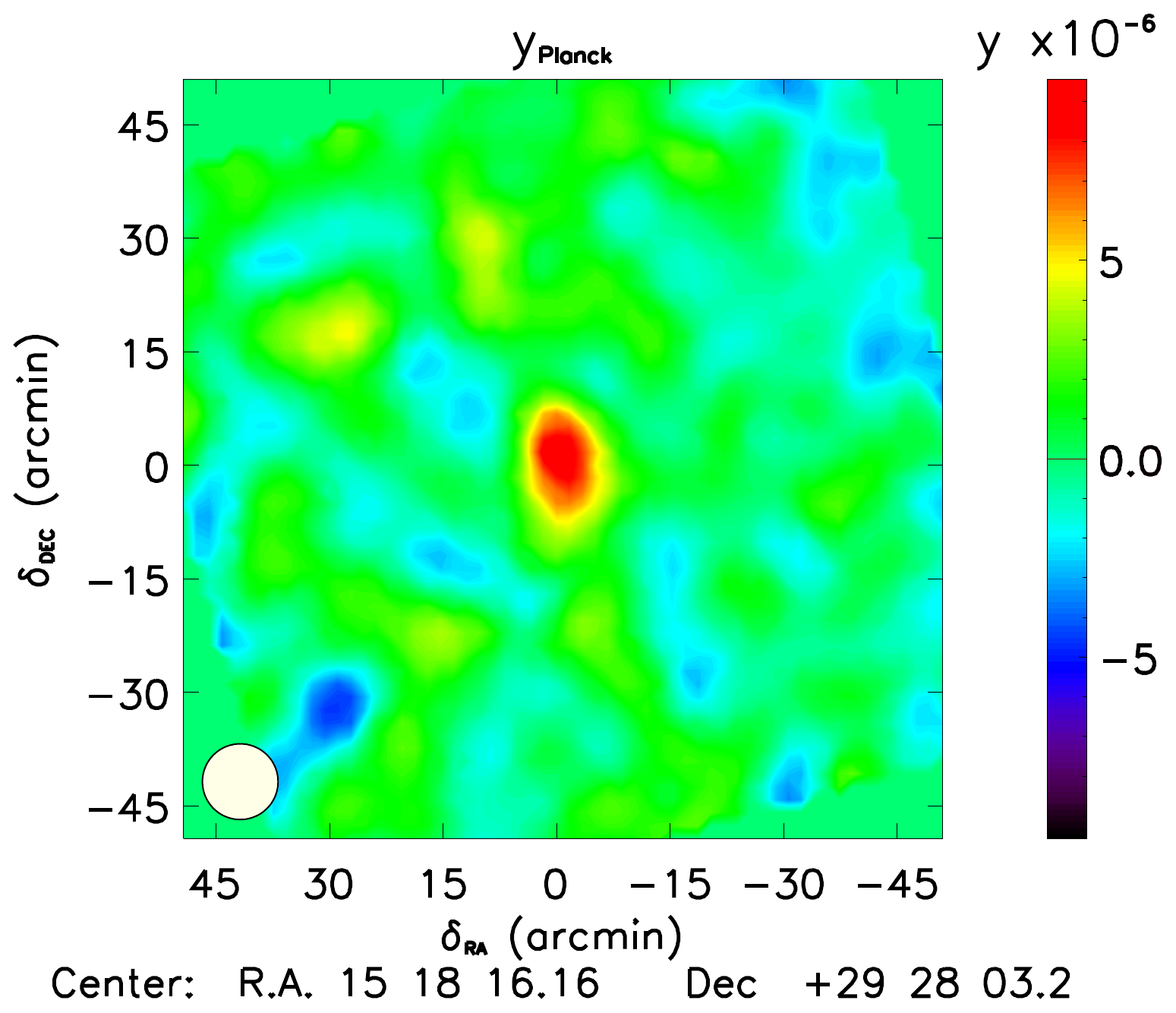}
\hspace{0.8cm}
\includegraphics[height=4.4cm]{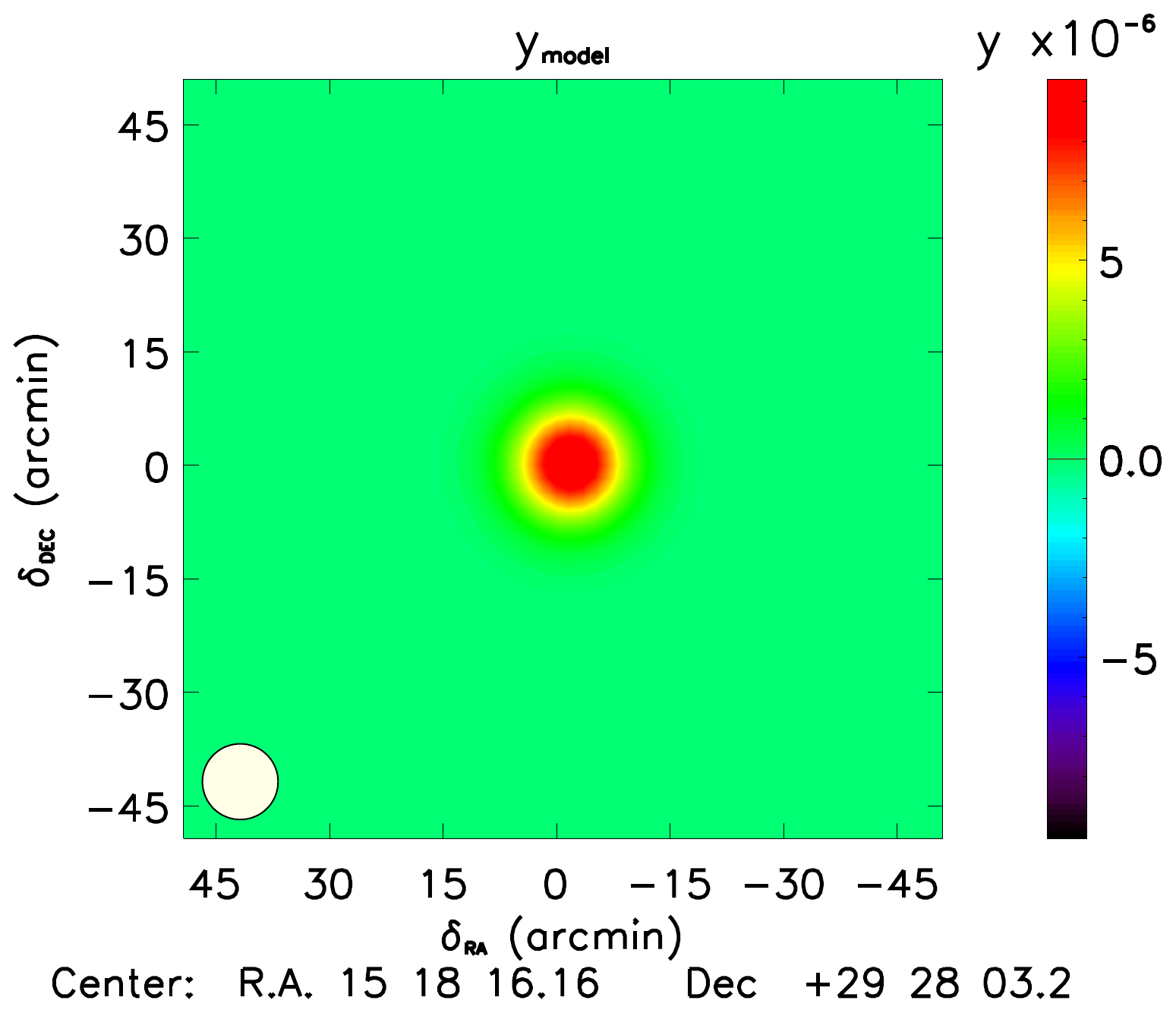}
\hspace{0.8cm}
\includegraphics[height=4.4cm]{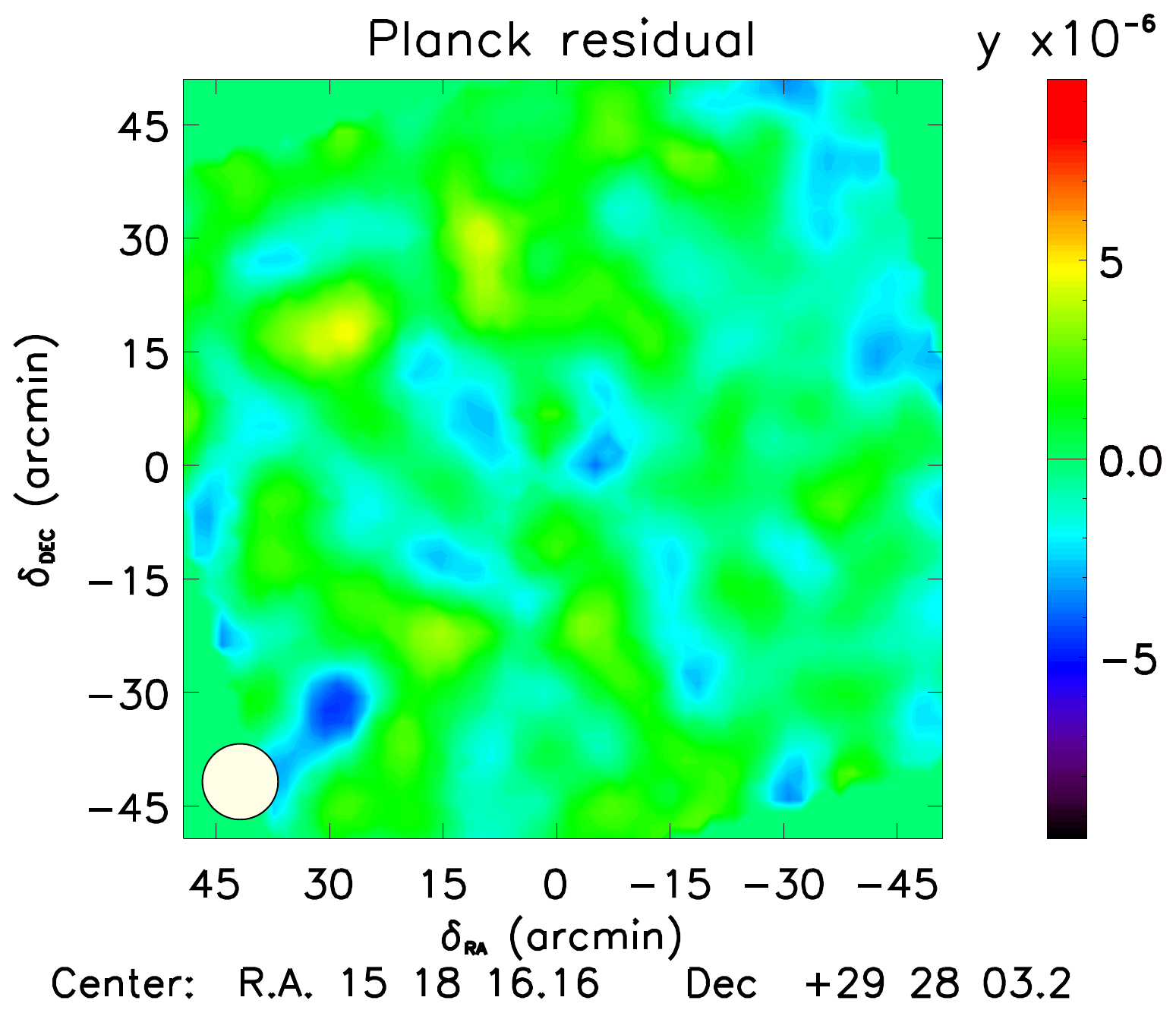}
\caption{{\footnotesize NIKA tSZ surface brightness at 150 GHz, maximum likelihood tSZ map and residual (top row) and \planck\ Compton parameter map, maximum likelihood Compton parameter map and residual (bottom row) computed from a non-parametric model based MCMC analysis. The residual maps does not indicate any significant substructure since no S/N over 3 is observed. The NIKA beam at 150 GHz and the Planck effective beam of 10 arcmin FWHM are shown in the bottom left-hand corner of the top and bottom row maps, respectively.}}
\label{fig:MCMC_maps}
\end{figure*}

The previous MCMC analysis has been upgraded to fully constrain the ICM pressure distribution from the cluster core to its outskirts and to improve on the integrated Compton parameter estimation. In this section, we present the new model that constrains the ICM pressure distribution, the new likelihood function used in the MCMC procedure, and the results obtained with this method.\\
\indent We use a non-parametric model to constrain the cluster pressure distribution in the framework of spherical symmetry to study potential deviations from the standard self-similar assumption (\emph{e.g.,} \citealt{NonparamPressure}). Instead of constraining the gNFW model parameters in the MCMC sampling, we constrain the values of the pressure at different distances from the X-ray center from the cluster core to its outskirts. The pressure between the constrained points is defined with a power law interpolation. We allow $P(r_i)$ and $P(r_{i+1})$ to be the constrained pressure at the $r_i$ and $r_{i+1}$ radii from the X-ray center, where the pressure at a radius $r \in [r_i, \, r_{i+1}]$ is defined as
\begin{equation}
        P(r) = P(r_i)\times 10^{\alpha}~~~\mathrm{with}~~~ \alpha = \frac{\mathrm{log_{10}}\left(\frac{P(r_{i+1})}{P(r_i)}\right) \times \mathrm{log_{10}}\left(\frac{r}{r_i}\right)}{\mathrm{log_{10}}\left(\frac{r_{i+1}}{r_i}\right)}
\label{eq:interpolation}
.\end{equation}
The pressure profile radial bins were defined with an increased sampling of the pressure profile in the region mainly constrained by the NIKA map (see Sec. \ref{sec:MCMC_old_NIKA}). Eleven pressure profile radial bins are defined from $\sim\!0.02 \, \mathrm{R_{500}}$ to $\sim\!\mathrm{R_{500}}$, which are mainly constrained by the NIKA tSZ surface brightness map, and 3 bins at $3$, $4,$ and $\sim\!5 \, \mathrm{R_{500}}$, which are constrained by the \planck\ Compton parameter map.\\
\indent Indeed, instead of using the \planck/AMI estimation of the integrated Compton parameter to partially break the degeneracy between the map zero level and the pressure profile characteristic radius, the Compton parameter map of \mbox{PSZ1\,G045.85+57.71} (see Fig. \ref{fig:Planck_XMM} left panel) that is obtained with MILCA (\citealt{ymap_planck}), $y_{\mathrm{Planck}}$, is used in combination with the NIKA map, $M_{\mathrm{NIKA}}$, to simultaneously constrain the cluster pressure profile at intermediate and large angular scales.\\
\indent We simulated \planck\ noise maps using the variance per pixel and homogeneous noise power spectrum provided by \citealt{ymap_planck}. These simulations are used to compute the pixel-to-pixel noise covariance matrix in the considered region of the \planck\ MILCA $y$-map $C_{\mathrm{Planck}}$. At each step of the MCMC sampling, a pressure profile is defined using equation \ref{eq:interpolation} and is used to derive a tSZ surface brightness map, $M_{\mathrm{model}}$ at 150~GHz, and a Compton parameter map $y_{\mathrm{model}}$. They are then respectively compared to the NIKA and \planck\ data via the following likelihood function:
\begin{equation}
\begin{tabular}{rl}
        $-2 \mathrm{ln} \, \mathscr{L}$  & $ =\chi^2_{\mathrm{NIKA}} + \chi^2_{\mathrm{Planck}}$\\[0.2cm]
         &$ =\sum_{i=1}^{N_{\mathrm{pixels}}^{\mathrm{NIKA}}} [(M_{\mathrm{NIKA}} - M_{\mathrm{model}})^T C_{\mathrm{NIKA}}^{-1} (M_{\mathrm{NIKA}} - M_{\mathrm{model}})]_i $ \\[0.2cm]
         & $+ \sum_{j=1}^{N_{\mathrm{pixels}}^{\mathrm{Planck}}} [(y_{\mathrm{Planck}} - y_{\mathrm{model}})^T C_{\mathrm{Planck}}^{-1} (y_{\mathrm{Planck}} - y_{\mathrm{model}})]_j$
\label{eq:chi2_NIKA_Planck}
\end{tabular}
.\end{equation}
Uniform priors spanning from 0 to $2~\mathrm{keV.cm^{-3}}$ are used for each pressure bin. This MCMC procedure also marginalizes over the zero level of the NIKA map and the calibration coefficient uncertainty. The correlations between the constrained pressure points are taken into account as in Sec. \ref{sec:MCMC_old_NIKA} to estimate the error bars on the pressure profile.\\
\indent We tested this method on simulations to check the pressure profile reconstruction. An input pressure distribution modeled as a gNFW profile was used to simulate tSZ surface brightness maps on which residual correlated noise was added using the noise power spectrum derived from the NIKA null-maps. The constrained pressure points are always consistent with the input pressure profile.\\
\indent The NIKA+\planck\ deprojected pressure profile of \mbox{PSZ1\,G045.85+57.71} is shown in Fig. \ref{fig:Pressure_XMM_NIKA_Planck} along with the XMM-{\it Newton} estimate in red. The pressure within the ICM is constrained from the cluster core to its outskirts without relying on X-ray spectroscopy at the intermediate redshift $z=0.61$. Such a non-parametric pressure profile deprojection is comparable with what has been achieved with the \planck\ satellite for low redshift ($z < 0.2$) galaxy clusters (see \citealt{Planck_pressure_prof}).\\
\indent The $1\sigma$ error bars of the deprojected pressure points are larger than the error bars we get from the previous MCMC analysis because the pressure profile is only constrained by the data whereas a parametric model fitting induces additional constraints. In the context of the spherical cluster assumption, such a non-parametric pressure profile deprojection gives an estimate of the intrinsic ICM pressure distribution without model-induced bias.\\
\indent Considering the \planck/AMI integrated Compton parameter in the MCMC analysis enables us to avoid models that diverge at large scales but does not takeall the information contained in the \planck\ Compton parameter map  into account. Using the whole \planck\ Compton parameter map in the likelihood estimation allows us to constrain both the normalization of the pressure profile and the pressure distribution at large scales where NIKA is not sensitive. Therefore, the three pressure profile radial bins constrained by the \planck\ map at large radii give a strong constraint on the pressure profile slope in the cluster outskirts. This highlights the complementarity between large FOV experiments, albeit with low resolution, such as \planck\ and the NIKA instrument, which benefit from the IRAM 30~m telescope high resolution.\\
\indent All the deprojected pressure values in Fig. \ref{fig:Pressure_XMM_NIKA_Planck} are compatible with the pressure profile derived with the previous MCMC analysis based on a gNFW modeling of the pressure distribution (see Sec. \ref{sec:MCMC_old_NIKA}). The pressure profile of \mbox{PSZ1\,G045.85+57.71} is therefore well-described by a gNFW model. The agreement between NIKA/\planck\ and XMM-{\it Newton} estimates is good as detailed in the following section.\\
\indent The maximum likelihood tSZ surface brightness map and Compton parameter map have been used to compute residual maps for both NIKA and \planck\ observations.  The top and bottom panels of Fig. \ref{fig:MCMC_maps} show the raw data, maximum likelihood model, and residual maps for NIKA and \planck, respectively. Although residuals are seen in the southwest region of the NIKA map, the S/N in both residual maps is always lower than 3, which therefore allows us to conclude that there are no significant substructures in \mbox{PSZ1\,G045.85+57.71} and that the NIKA 150 GHz map of this cluster is well described by a projected spherical model, given the amount of residual correlated noise.
\begin{table*}[h]
\begin{center}
\begin{tabular}{ccc}
\hline
\hline
Data & Method & $Y_{500}$ ($\mathrm{arcmin^2}$) \\
\hline
\planck\ & catalog &$8.21^{+1.73}_{-1.70} \times 10^{-4}$\\
NIKA + \planck/AMI & parametric &$4.23^{+0.68}_{-0.62} \times 10^{-4}$\\
NIKA + \planck\ map & non-parametric & $5.61^{+0.68}_{-0.59} \times 10^{-4}$\\
\hline
\hline
\end{tabular}
\end{center}
\caption{{\footnotesize Estimations of \mbox{PSZ1\,G045.85+57.71} integrated Compton parameter ($\mathrm{Y_{500}}$) from the constraint derived by \planck\ (\citealt{PSZ1_updated}), from the 2D model based MCMC analysis (see Sec. \ref{sec:MCMC_old}), and from the non-parametric model based MCMC analysis (see Sec. \ref{sec:nomodeling}).}}
\label{tab:Cluster_global}
\end{table*}

\subsection{Integrated Compton parameter estimation}\label{sec:int_compton}

Both parametric and non-parametric deprojected pressure profiles are then used to estimate the cluster integrated Compton parameter \mbox{$\rm{Y_{500}^{param}} = 4.23^{+0.68}_{-0.62} \times 10^{-4} \, \mathrm{arcmin^2}$} and \mbox{$\rm{Y_{500}^{non-param}} = 5.61^{+0.68}_{-0.59} \times 10^{-4} \, \rm{arcmin^2}$}, which are in agreement with the \planck\ estimation \mbox{$\rm{Y_{500}} = 8.21^{+1.73}_{-1.70} \times 10^{-4} \, \mathrm{arcmin^2}$} (\citealt{PSZ1_updated}). The best relative uncertainty is obtained with the non-parametric pressure profile deprojection because this method gives the most stringent constraints on the cluster pressure distribution from its core up to $5\mathrm{R_{500}}$. Thus, the relative uncertainty on the integrated Compton parameter tracing the total thermal energy within the ICM is improved by a factor 2 with respect to the \planck\ estimate because the pressure profile is much more constrained at each scale. Furthermore, the NIKA high angular resolution allows us to completely break the $\theta_s - \rm{Y_{tot}}$ degeneracy observed in both \planck\ and AMI observations (\citealt{AMI_followup}).\\
\indent This result highlights the necessity of a high resolution tSZ follow-up of \planck-discovered clusters to better constrain the $Y-M$ scaling relation used for future cosmology studies. Indeed, a non-parametric joint analysis of both NIKA and \planck\ data leads to a deprojected pressure profile, which is constrained at every scale and is not affected by model-induced bias. This approach allows us to give a stringent constraint on the integrated Compton parameter used to calibrate the $Y-M$ scaling relation. The \mbox{PSZ1\,G045.85+57.71} integrated Compton parameter estimations derived from SZ observations are summarized in Table \ref{tab:Cluster_global}.

%========== Results
\subsection{Thermodynamics of the cluster}\label{sec:comparison_X_tSZ}
\begin{figure*}[h]
\centering
\includegraphics[height=5.5cm]{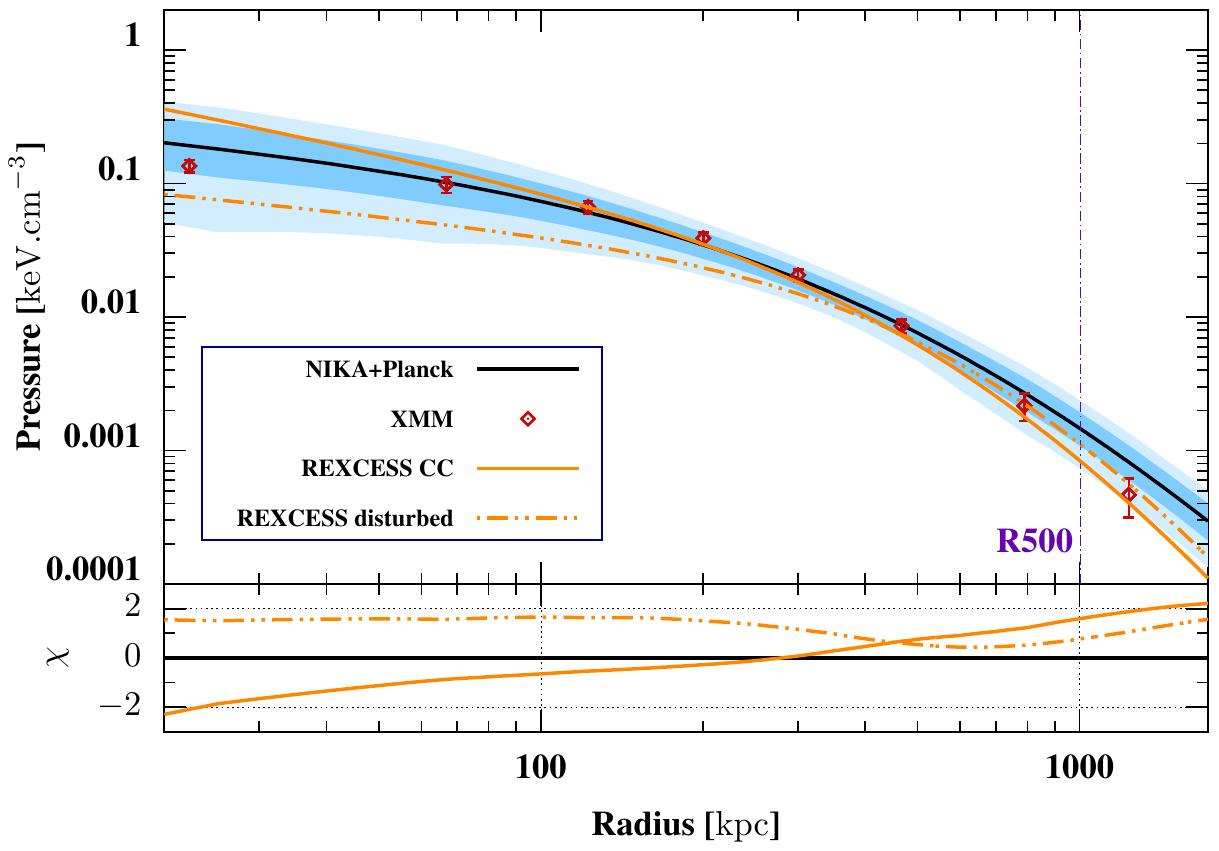}
\hspace{1.3cm}
\includegraphics[height=5.5cm]{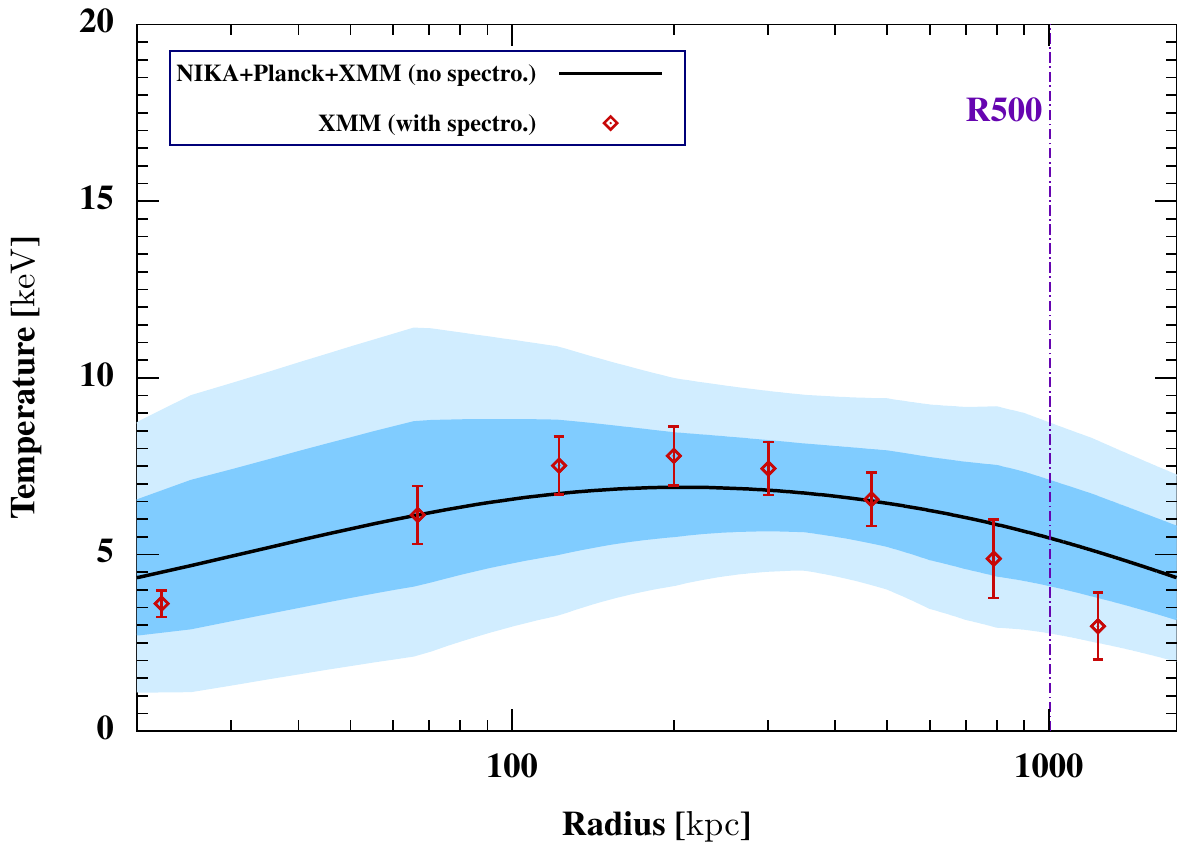}
\includegraphics[height=5.5cm]{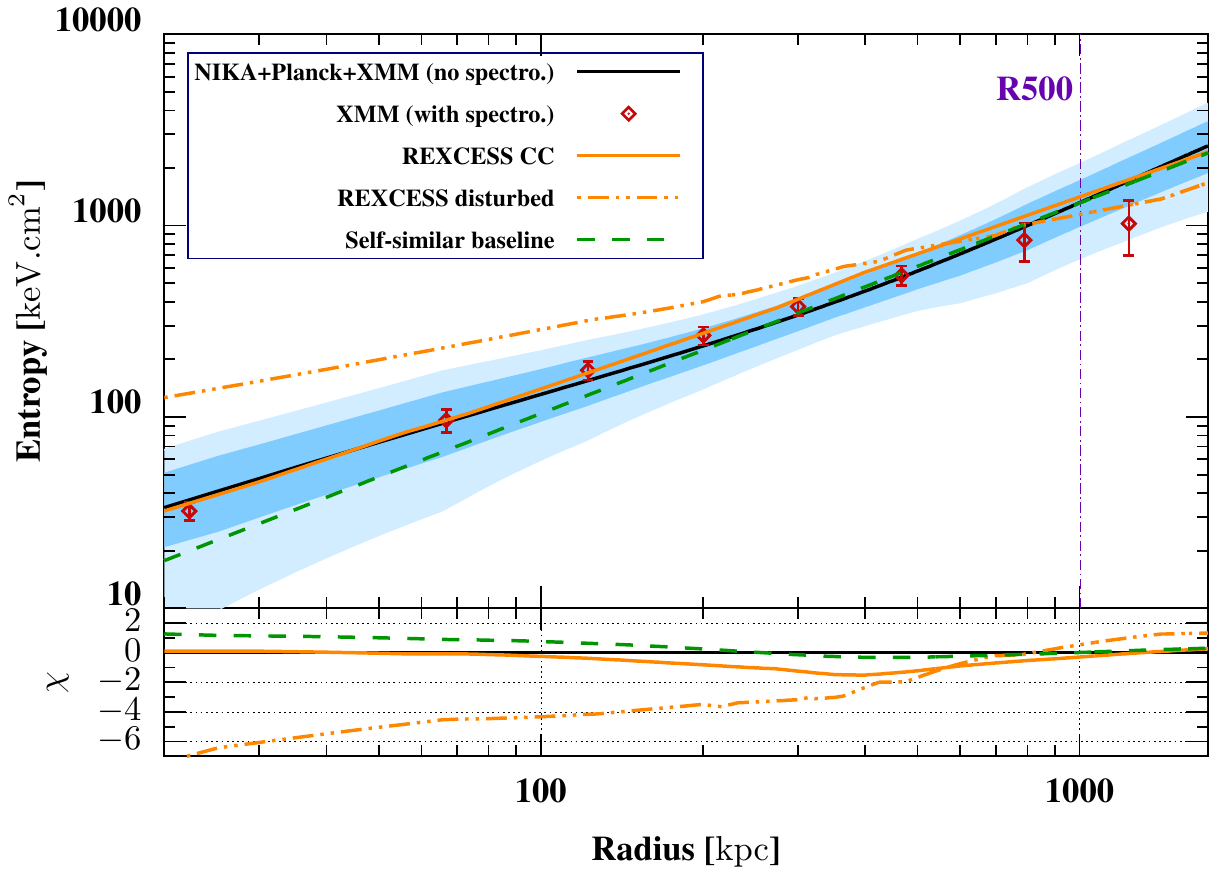}
\hspace{1.3cm}
\includegraphics[height=5.7cm]{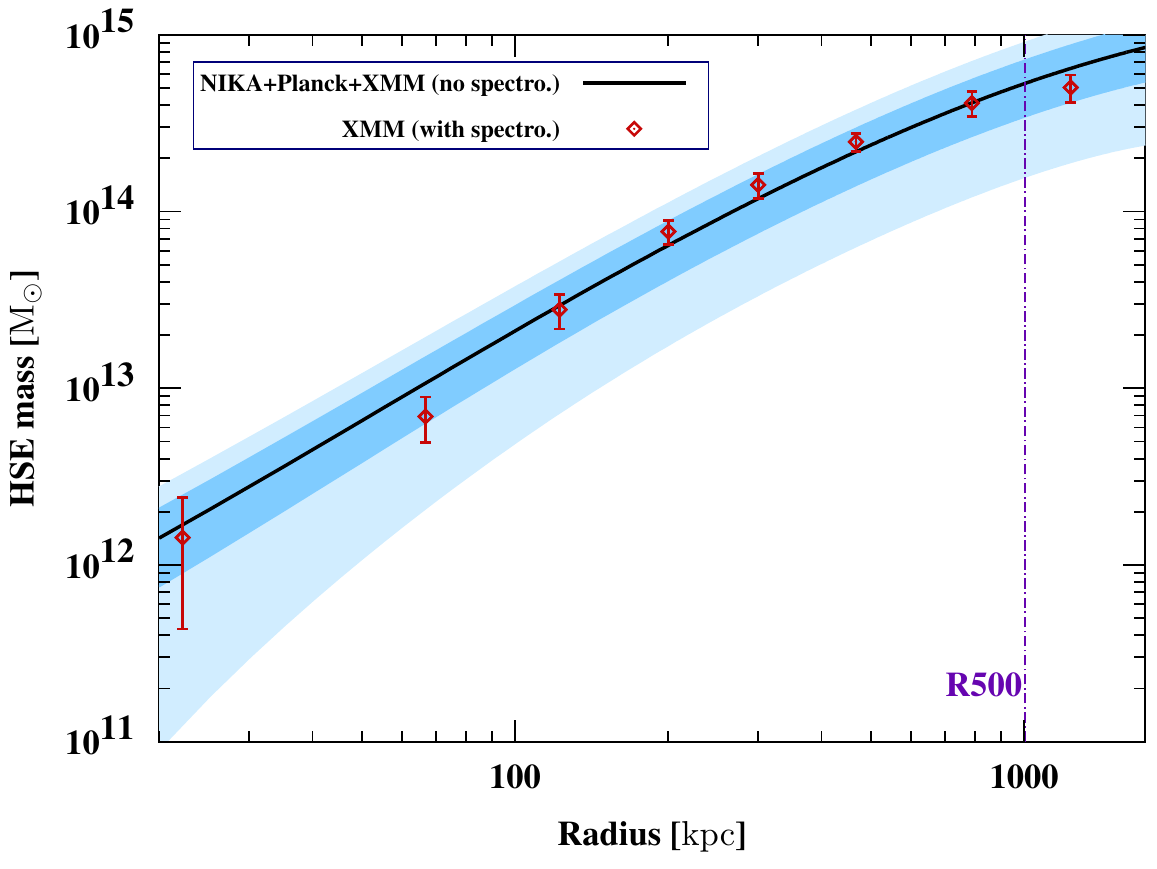}
\caption{{\footnotesize MCMC constraints on the deprojected radial profiles of the pressure (top left), temperature (top right), entropy (bottom left), and hydrostatic mass (bottom right) based on the non-parametric method. The XMM-{\it Newton} only measurements are indicated with red dots. The dark and light blue regions show the 68\% and 95\% confidence limit on the NIKA/\planck\ estimated profiles, whose best fit is indicated by the black line. The pressure (\citealt{universal}) and entropy (\citealt{entropy_REXCESS}) mean profiles of both cool-core (orange solid line) and morphologically disturbed (orange dashed line) clusters based on a representative sample of nearby X-ray clusters is also shown. The weighted difference between these mean profiles and the NIKA/\planck\ estimated profile is shown in the bottom part of both pressure and entropy panels. For the entropy profile, the self-similar expectation computed from non-radiative simulations (\citealt{entropy_base}) is also represented as a green dashed line.}}
\label{fig:thermo_NIKA}
\end{figure*}
In this section, we use the complementarity between NIKA and XMM-{\it Newton} data sets to fully constrain the thermodynamics of \mbox{PSZ1\,G045.85+57.71}. As both the cluster density and pressure profiles estimated from non-parametric deprojection methods are consistent with parametric models (see Fig. \ref{fig:XMM_density} and \ref{fig:Pressure_XMM_NIKA_Planck}), we choose for convenience to combine the best-fit SVM model of the XMM-{\it Newton} density profile and the best-fit gNFW model of the NIKA/\planck\ pressure profile to constrain the whole ICM thermodynamics.\\ 
%---------- Pressure profile
\indent The maximum likelihood pressure values computed from the non-parametric analysis were therefore fitted by a gNFW model by taking into account the correlations between each pressure points. A reduced $\chi^2$ of $1.13$ was computed for the fit. This emphasizes the good agreement between the NIKA deprojected pressure points and the standard gNFW model. The pressure profile constrained by the 150 GHz NIKA map is shown in Fig. \ref{fig:thermo_NIKA} (top left panel) in black together with the deprojected pressure points from the XMM-{\it Newton} analysis in red. All the XMM-{\it Newton} estimated pressure points are compatible with the NIKA constrained pressure profile within the 68\% confidence level uncertainties shown in blue in Fig. \ref{fig:thermo_NIKA}. The XMM-{\it Newton} estimate of the pressure profile can only be inferred with spectroscopic information while tSZ observations directly probe the pressure distribution within the ICM. Comparing both estimated pressure profiles allows us then to bring strong constraints on the cluster pressure distribution as the two methods are completely independent.\\
\indent The NIKA pressure profile estimate is compared with the universal pressure profile computed using the \rexcess\ representative sample of nearby clusters (\citealt{boe07,universal}). The solid and dashed orange lines in Fig. \ref{fig:thermo_NIKA} (top left panel) give the cool-core and morphologically disturbed subsample mean pressure profile, respectively. The normalization of the two profiles was computed using the XMM-{\it Newton} total mass estimation taking into account the mass dependence of the shape of the profile (\citealt{universal}). As shown in the bottom part of Fig. \ref{fig:thermo_NIKA} (top left panel), the cool-core and morphologically disturbed cluster universal profiles are both within the $2\sigma$ error bars of the \mbox{PSZ1\,G045.85+57.71} pressure profile estimation. Therefore, the NIKA estimated profile alone does not bring significant information on the relaxation state of this cluster.\\
%---------- Entropy and temperature (X+spectro Vs combined tSZ+Xray)
\indent The cluster temperature, entropy, and mass profiles were computed by combining both the NIKA estimated pressure profile and the fitted SVM density profile as explained in Sec. \ref{sec:modeling}.\\
\indent The estimated temperature profile shown in Fig. \ref{fig:thermo_NIKA} (top right panel) is compatible with that estimated from the XMM-{\it Newton} spectroscopic observations and its shape is consistent with that expected for a cool-core cluster. The core temperature goes down to $\sim$4~keV and the maximum temperature of  $\sim$7~keV is reached at a distance of $\sim$200~kpc from the X-ray center. The NIKA-XMM combined temperature profile (without spectroscopy) seems to be flatter in the cluster outskirts than that estimated by the XMM data alone (with spectroscopy). This could be an indication of clumping in the cluster outskirts. However, this trend is not significant compared to the error bars.\\
\indent The estimated entropy profile is shown in Fig. \ref{fig:thermo_NIKA} (bottom left panel) along with the XMM-{\it Newton} results. As shown in \citealt{Voit_cosmo}, the entropy distribution in the ICM traces the thermodynamical history of the gas and is a good estimator of its relaxation state. A baseline entropy profile was computed by \citealt{entropy_base} using numerical simulations without including hydrodynamical processes. The latter, converted from an overdensity of 200 to 500, takes the form of a power law scaled by a factor depending on the cluster mass and baryon mass fraction (\citealt{entropy_REXCESS}), 
\begin{equation}
K(r) = 1.42  \, K_{\mathrm{500}} \, (R/R_{\mathrm{500}})^{1.1}~~\mathrm{with}
\label{eq:baseline_entropy}
\end{equation}
$$K_{\mathrm{500}} = 106~\mathrm{keV \, cm^{-2}} \left(\frac{M_{\mathrm{500}}}{10^{14}h_{\mathrm{70}}^{-1}M_{\odot}}\right)^{2/3} \, \left(\frac{1}{f_b}\right)^{2/3} \, E(z)^{-2/3} \, h_{\mathrm{70}}^{-4/3}.$$
The corresponding self-similar baseline was computed for this cluster and is shown as a green dashed line in Fig. \ref{fig:thermo_NIKA} (bottom left panel). The concordance between this baseline and the estimated entropy profile is very good especially outside the cluster core where the non-gravitational processes have less impact on the derived thermodynamic constraints. In order to have a better description of the entropy distribution in both the cluster core and its periphery, it can be modeled by a power law plus constant profile (\citealt{entropy_constant}),
\begin{equation}
K(r) = K_0 + K_{100}\left(\frac{r}{100h^{-1}_{70}~\mathrm{kpc}}\right)^{\alpha}
\label{eq:entro_REXCESS}
.\end{equation}
This model describes well the higher plateau and shallower slope observed on disturbed system entropy profiles. The NIKA-XMM combined entropy profile is compared with the mean entropy profiles estimated from the \rexcess\ representative subsamples of cool-core and morphologically disturbed clusters using the best-fit estimations of these model parameters (\citealt{entropy_REXCESS}). The results are shown in Fig. \ref{fig:thermo_NIKA} (bottom left panel) using a solid line and a dashed orange line for the cool-core and morphologically disturbed clusters, respectively. The bottom part of the figure shows the weighted difference between the NIKA-XMM estimated entropy profile and the considered models. The mean entropy profile computed from the \rexcess\ representative subsample of cool-core clusters is in very good agreement with the NIKA-XMM estimated profile especially in the cluster core. The mean profile describing the morphologically disturbed cluster entropy distribution is however in strong tension with the observed profile. Indeed, the deviation from the NIKA-XMM estimated profile is higher than $3\sigma$ from the X-ray center up to radial scales of $\sim$400~kpc. The estimated entropy profile allows us to conclude that \mbox{PSZ1\,G045.85+57.71} is a cool-core cluster, confirming the indications from the temperature profile. This emphasizes the complementarity between tSZ and X-ray observations to constrain the full thermodynamic state of a cluster.\\ 
%---------- Mass and Ytot
\indent The hydrostatic equilibrium hypothesis was assumed to derive the mass profile of \mbox{PSZ1\,G045.85+57.71} as described in Sec. \ref{sec:modeling}. The estimated profile is shown in the bottom right panel of Fig. \ref{fig:thermo_NIKA} along with the XMM-{\it Newton} constraints using only X-ray data. All the XMM-{\it Newton} estimated values are compatible with the NIKA-XMM combined profile within the 68\% confidence limit. The estimated mass profile was then used to compute the cluster characteristic radius \mbox{$\rm{R_{500}} = 1004^{+202}_{-161}~kpc$} and total mass within $\rm{R_{500}}$, \mbox{$\rm{M_{500}} = (5.4^{+2.6}_{-3.0})\times 10^{14} \, \rm{M_{\odot}}$}. We do not expect to obtain constraints as stringent as those derived from an X-ray based analysis because the reconstructed NIKA pressure profile shows a larger dispersion. Nevertheless, these results are compatible with the XMM-{\it Newton} estimations using spectroscopic observations: \mbox{$\rm{R_{500}^{XMM}} = 1013 \pm 13~kpc$} and \mbox{$\rm{M_{500}^{XMM}} = (5.78 \pm 0.21)\times 10^{14} \, \rm{M_{\odot}}$} and show that tSZ observations are a good alternative to derive cluster thermodynamic properties even at high redshift, where accurate X-ray spectroscopy measurements require large integration time.

%##############################################################################
%##########################                 CONCLUSION                ##########################%##############################################################################
\section{Conclusions and perspectives}\label{sec:conclusions}
%---------- Short review
The \planck\ tSZ-discovered cluster \psz\ has been observed simultaneously at 150 and 260~GHz by the NIKA camera. A 4.35 hour observation allowed a detailed mapping at 18.2 arcsec angular resolution of the tSZ signal at 150~GHz. The cluster was also observed in the X-ray band by the \xmm\ satellite.

%---------- Non-parametric pressure profile deprojection
We performed the first non-parametric pressure profile deprojection from resolved tSZ observations of a \planck-discovered cluster at an intermediate redshift ($z=0.61$). The MCMC procedure, which was developed to deproject the cluster pressure profile, uses the NIKA tSZ surface brightness map and the \planck\ Compton parameter map jointly to constrain the cluster pressure distribution from its core up to $5R_{500}$. The resulting pressure profile does not deviate significantly from the standard gNFW model.

%---------- NIKA/Planck complementarity
The combination of both NIKA and \planck\ data brings strong constraints on the pressure profile slope at each scale, and allows a significant improvement in the relative uncertainty on the integrated Compton parameter value $\Yv$. The latter highlights the utility of high resolution tSZ follow-up of \planck-discovered clusters to better constrain the $Y$--$M$ scaling relation used for cosmology studies based on cluster counts \citep{ProceedingMoriond}.

%--------- NIKA/XMM complementarity 
We further combined the NIKA+\planck\ deprojected non-parametric pressure profile with the deprojected electronic density profile obtained from \xmm\ observations. This allowed us to obtain temperature and entropy profiles without recourse to X-ray spectroscopy and to undertake an hydrostatic mass analysis. The X-ray only (including spectroscopy) and the tSZ$+$X-ray (without spectroscopy) constraints are consistent within their uncertainties. This shows that high resolution tSZ observations, combined with X-ray snapshot imagery, are a competitive alternative to constrain cluster thermodynamics at high redshift, where X-ray spectroscopy requires large integration times to derive accurate temperature estimates. Comparison of the thermodynamic profiles to those obtained from the representative X-ray sample \rexcess\ \citep{boe07,universal,entropy_REXCESS}, in particular the radial distributions of temperature and entropy, indicates that \psz\ is a cool-core cluster. This result illustrates the complementarity between tSZ and X-ray data when only X-ray imaging observations are available.\\

%---------- Prospect for NIKA2
The NIKA2 camera now installed at the focal plane of the IRAM 30-m telescope is currently undergoing commissioning. The number of detectors has been increased by a factor 10 with respect to the NIKA prototype to fully sample the telescope field of view of 6.5~arcmin. The NIKA2 tSZ Guaranteed Time Large Program \citep{NIKA2LP} is a follow-up of 50 SZ-discovered clusters with redshift up to $z = 1$, selected from the \planck\ and ACT catalogs \citep{ACT_cluster,Planck_cata2}. Following the work presented in this paper and in the previous NIKA studies \citep{RXJ1347NIKA,CLJ1227NIKA,MACSJ1424NIKA}, NIKA2 is expected to provide reliable tSZ detection and mapping of galaxy clusters in only a few hours integration time per cluster. Although NIKA2 alone will be a key tool for further understanding cluster physics, using the complementarity between different observational probes constitutes the best road for getting a comprehensive picture of the ICM. The NIKA2 data will therefore be complemented with ancillary data including X-ray, optical, and radio observations. The full data set will lead to significant improvements on the use of galaxy clusters to obtain constraints on cosmology and on the matter distribution and content of the Universe.

%###############################################################################################
%##########################                       ACKNOWLEDGEMENTS                        ##########################%###############################################################################################
\begin{acknowledgements}
We would like to thank the IRAM staff for their support during the campaigns. 
The NIKA dilution cryostat has been designed and built at the Institut N\'eel. 
In particular, we acknowledge the crucial contribution of the Cryogenics Group, and 
in particular Gregory Garde, Henri Rodenas, Jean Paul Leggeri, Philippe Camus. 
This work has been partially funded by the Foundation Nanoscience Grenoble, the LabEx FOCUS ANR-11-LABX-0013, and 
the ANR under the contracts "MKIDS", "NIKA", and ANR-15-CE31-0017. 
This work has benefited from the support of the European Research Council Advanced Grant ORISTARS 
under the European Union's Seventh Framework Programme (Grant Agreement no. 291294).
We acknowledge fundings from the ENIGMASS French LabEx (R. A. and F. R.), 
the CNES post-doctoral fellowship program (R. A.),  the CNES doctoral fellowship program (A. R.), and 
the FOCUS French LabEx doctoral fellowship program (A. R.).
\end{acknowledgements}


\begin{thebibliography}{81}
\expandafter\ifx\csname natexlab\endcsname\relax\def\natexlab#1{#1}\fi

\bibitem[{Adam(2015)}]{these_remi}
Adam, R. 2015, PhD thesis, Universite Grenoble Alpes, France

\bibitem[{Adam {et~al.}(2014)}]{RXJ1347NIKA}
Adam, R., {et~al.} 2014, \aap, 569, A66, 1310.6237

\bibitem[{Adam {et~al.}(2015)}]{CLJ1227NIKA}
------. 2015, \aap, 576, A12, 1410.2808

\bibitem[{Adam {et~al.}(2016{\natexlab{a}})}]{MACSJ1424NIKA}
------. 2016{\natexlab{a}}, \aap, 586, A122, 1510.06674

\bibitem[{Adam {et~al.}(2016{\natexlab{b}})}]{MACSJ0717NIKA}
------. 2016{\natexlab{b}}, 1606.07721

\bibitem[{Ameglio {et~al.}(2007)Ameglio, Borgani, Pierpaoli, \&
  Dolag}]{NonparamSimu3}
Ameglio, S., Borgani, S., Pierpaoli, E., \& Dolag, K. 2007, \mnras, 382, 397

\bibitem[{{Anderson} {et~al.}(2014){Anderson}, {Aubourg}, {Bailey}, {Beutler},
  {Bhardwaj}, {Blanton}, {Bolton}, {Brinkmann}, {Brownstein}, {Burden},
  {Chuang}, {Cuesta}, {Dawson}, {Eisenstein}, {Escoffier}, {Gunn}, {Guo}, {Ho},
  {Honscheid}, {Howlett}, {Kirkby}, {Lupton}, {Manera}, {Maraston}, {McBride},
  {Mena}, {Montesano}, {Nichol}, {Nuza}, {Olmstead}, {Padmanabhan},
  {Palanque-Delabrouille}, {Parejko}, {Percival}, {Petitjean}, {Prada},
  {Price-Whelan}, {Reid}, {Roe}, {Ross}, {Ross}, {Sabiu}, {Saito}, {Samushia},
  {S{\'a}nchez}, {Schlegel}, {Schneider}, {Scoccola}, {Seo}, {Skibba},
  {Strauss}, {Swanson}, {Thomas}, {Tinker}, {Tojeiro}, {Maga{\~n}a}, {Verde},
  {Wake}, {Weaver}, {Weinberg}, {White}, {Xu}, {Y{\`e}che}, {Zehavi}, \&
  {Zhao}}]{and14}
{Anderson}, L. {et~al.} 2014, \mnras, 441, 24, 1312.4877

\bibitem[{{Applegate} {et~al.}(2014){Applegate}, {von der Linden}, {Kelly},
  {Allen}, {Allen}, {Burchat}, {Burke}, {Ebeling}, {Mantz}, \&
  {Morris}}]{app14}
{Applegate}, D.~E. {et~al.} 2014, \mnras, 439, 48, 1208.0605

\bibitem[{Arnaud {et~al.}(2010)Arnaud, Pratt, Piffaretti, Boehringer, Croston,
  \& Pointecouteau}]{universal}
Arnaud, M., Pratt, G.~W., Piffaretti, R., Boehringer, H., Croston, J.~H., \&
  Pointecouteau, E. 2010, \aap, 517, A92, 0910.1234

\bibitem[{Basu {et~al.}(2010)}]{NonparamPressure}
Basu, K., {et~al.} 2010, \aap, 519, A29, 0911.3905

\bibitem[{Becker {et~al.}(1995)Becker, White, \& Helfand}]{Becker1995}
Becker, R.~H., White, R.~L., \& Helfand, D.~J. 1995, \apj, 450, 559

\bibitem[{Birkinshaw(1999)}]{Birkinshaw}
Birkinshaw, M. 1999, Phys. Rept., 310, 97, astro-ph/9808050

\bibitem[{{Biviano} {et~al.}(2006){Biviano}, {Murante}, {Borgani}, {Diaferio},
  {Dolag}, \& {Girardi}}]{biv06}
{Biviano}, A., {Murante}, G., {Borgani}, S., {Diaferio}, A., {Dolag}, K., \&
  {Girardi}, M. 2006, \aap, 456, 23, astro-ph/0605151

\bibitem[{Bleem {et~al.}(2015)}]{SPTcluster}
Bleem, L.~E., {et~al.} 2015, \apjs, 216, 27, 1409.0850

\bibitem[{{B{\"o}hringer} {et~al.}(2010){B{\"o}hringer}, {Pratt}, {Arnaud},
  {Borgani}, {Croston}, {Ponman}, {Ameglio}, {Temple}, \& {Dolag}}]{boe10}
{B{\"o}hringer}, H. {et~al.} 2010, \aap, 514, A32, 0912.4667

\bibitem[{{B{\"o}hringer} {et~al.}(2007){B{\"o}hringer}, {Schuecker}, {Pratt},
  {Arnaud}, {Ponman}, {Croston}, {Borgani}, {Bower}, {Briel}, {Collins},
  {Donahue}, {Forman}, {Finoguenov}, {Geller}, {Guzzo}, {Henry}, {Kneissl},
  {Mohr}, {Matsushita}, {Mullis}, {Ohashi}, {Pedersen}, {Pierini}, {Quintana},
  {Raychaudhury}, {Reiprich}, {Romer}, {Rosati}, {Sabirli}, {Temple}, {Viana},
  {Vikhlinin}, {Voit}, \& {Zhang}}]{boe07}
------. 2007, \aap, 469, 363, arXiv:astro-ph/0703553

\bibitem[{{Bonamente} {et~al.}(2012){Bonamente}, {Hasler}, {Bulbul},
  {Carlstrom}, {Culverhouse}, {Gralla}, {Greer}, {Hawkins}, {Hennessy}, {Joy},
  {Kolodziejczak}, {Lamb}, {Landry}, {Leitch}, {Marrone}, {Miller},
  {Mroczkowski}, {Muchovej}, {Plagge}, {Pryke}, {Sharp}, \& {Woody}}]{bon12}
{Bonamente}, M. {et~al.} 2012, New Journal of Physics, 14, 025010, 1112.1599

\bibitem[{Bourrion {et~al.}(2012)Bourrion, Vescovi, Bouly, Benoit, Calvo,
  Gallin-Martel, Macias-Perez, \& Monfardini}]{NIKA_elec2}
Bourrion, O., Vescovi, C., Bouly, J.~L., Benoit, A., Calvo, M., Gallin-Martel,
  L., Macias-Perez, J.~F., \& Monfardini, A. 2012, JINST, 7, P07014, 1204.1415

\bibitem[{{Calvo} {et~al.}(2013){Calvo}, {Roesch}, {D{\'e}sert}, {Monfardini},
  {Benoit}, {Mauskopf}, {Ade}, {Boudou}, {Bourrion}, {Camus}, {Cruciani},
  {Doyle}, {Hoffmann}, {Leclercq}, {Macias-Perez}, {Ponthieu}, {Schuster},
  {Tucker}, \& {Vescovi}}]{cal13}
{Calvo}, M. {et~al.} 2013, \aap, 551, L12

\bibitem[{Carlstrom {et~al.}(2002)Carlstrom, Holder, \& Reese}]{Carlstrom}
Carlstrom, J.~E., Holder, G.~P., \& Reese, E.~D. 2002, \araa, 40, 643,
  astro-ph/0208192

\bibitem[{Catalano {et~al.}(2014)}]{NIKA_calib}
Catalano, A., {et~al.} 2014, \aap, 569, A9

\bibitem[{Catalano {et~al.}(2016)}]{ProceedingSPIE}
------. 2016, 1605.08628

\bibitem[{Comis {et~al.}(2016)}]{ProceedingMoriond}
Comis, B., {et~al.} 2016, in {51st Rencontres de Moriond on Cosmology La
  Thuile, Italy, March 19-26, 2016}, 1605.09549

\bibitem[{Condon {et~al.}(1998)Condon, Cotton, Greisen, Yin, Perley, Taylor, \&
  Broderick}]{Condon1998}
Condon, J.~J., Cotton, W.~D., Greisen, E.~W., Yin, Q.~F., Perley, R.~A.,
  Taylor, G.~B., \& Broderick, J.~J. 1998, AJ, 115, 1693

\bibitem[{{Croston} {et~al.}(2006){Croston}, {Arnaud}, {Pointecouteau}, \&
  {Pratt}}]{cro06}
{Croston}, J.~H., {Arnaud}, M., {Pointecouteau}, E., \& {Pratt}, G.~W. 2006,
  \aap, 459, 1007, arXiv:astro-ph/0608700

\bibitem[{da~Silva {et~al.}(2004)da~Silva, Kay, Liddle, \& Thomas}]{das04}
da~Silva, A.~C., Kay, S.~T., Liddle, A.~R., \& Thomas, P.~A. 2004, \mnras, 348,
  1401

\bibitem[{{de Haan} {et~al.}(2016){de Haan}, {Benson}, {Bleem}, {Allen},
  {Applegate}, {Ashby}, {Bautz}, {Bayliss}, {Bocquet}, {Brodwin}, {Carlstrom},
  {Chang}, {Chiu}, {Cho}, {Clocchiatti}, {Crawford}, {Crites}, {Desai},
  {Dietrich}, {Dobbs}, {Doucouliagos}, {Foley}, {Forman}, {Garmire}, {George},
  {Gladders}, {Gonzalez}, {Gupta}, {Halverson}, {Hlavacek-Larrondo},
  {Hoekstra}, {Holder}, {Holzapfel}, {Hou}, {Hrubes}, {Huang}, {Jones},
  {Keisler}, {Knox}, {Lee}, {Leitch}, {von der Linden}, {Luong-Van}, {Mantz},
  {Marrone}, {McDonald}, {McMahon}, {Meyer}, {Mocanu}, {Mohr}, {Murray},
  {Padin}, {Pryke}, {Rapetti}, {Reichardt}, {Rest}, {Ruel}, {Ruhl},
  {Saliwanchik}, {Saro}, {Sayre}, {Schaffer}, {Schrabback}, {Shirokoff},
  {Song}, {Spieler}, {Stalder}, {Stanford}, {Staniszewski}, {Stark}, {Story},
  {Stubbs}, {Vanderlinde}, {Vieira}, {Vikhlinin}, {Williamson}, \&
  {Zenteno}}]{deh16}
{de Haan}, T. {et~al.} 2016, 1603.06522

\bibitem[{{D{\'e}mocl{\`e}s} {et~al.}(2010){D{\'e}mocl{\`e}s}, {Pratt},
  {Pierini}, {Arnaud}, {Zibetti}, \& {D'Onghia}}]{dem10}
{D{\'e}mocl{\`e}s}, J., {Pratt}, G.~W., {Pierini}, D., {Arnaud}, M., {Zibetti},
  S., \& {D'Onghia}, E. 2010, \aap, 517, A52, 1005.0320

\bibitem[{Donahue {et~al.}(2006)Donahue, Horner, Cavagnolo, \&
  Voit}]{entropy_constant}
Donahue, M., Horner, D.~J., Cavagnolo, K.~W., \& Voit, G.~M. 2006, \apj, 643,
  730, astro-ph/0511401

\bibitem[{{Ebeling}(2014)}]{HST_followup}
{Ebeling}, H. 2014, {Beyond MACS: A Snapshot Survey of the Most Massive
  Clusters of Galaxies at z $>$ 0.5}, HST Proposal

\bibitem[{{Eckert} {et~al.}(2013){Eckert}, {Ettori}, {Molendi}, {Vazza}, \&
  {Paltani}}]{eck13}
{Eckert}, D., {Ettori}, S., {Molendi}, S., {Vazza}, F., \& {Paltani}, S. 2013,
  \aap, 551, A23, 1301.0624

\bibitem[{Gavazzi(2005)}]{Projection_effect}
Gavazzi, R. 2005, \aap, 443, 793, astro-ph/0503696

\bibitem[{Gelman \& Rubin(1992)}]{convergence_MCMC}
Gelman, A., \& Rubin, D.~B. 1992, Statist. Sci., 7, 457

\bibitem[{Griffin {et~al.}(2010)}]{SPIRE}
Griffin, M.~J., {et~al.} 2010, \aap, 518, L3, 1005.5123

\bibitem[{Halverson {et~al.}(2009)Halverson, Lanting, {et~al.}}]{NonparamApex}
Halverson, N., Lanting, T., {et~al.} 2009, \apj, 701, 42

\bibitem[{Hasselfield {et~al.}(2013)}]{ACT_cluster}
Hasselfield, M., {et~al.} 2013, JCAP, 1307, 008, 1301.0816

\bibitem[{{Hoekstra} {et~al.}(2015){Hoekstra}, {Herbonnet}, {Muzzin}, {Babul},
  {Mahdavi}, {Viola}, \& {Cacciato}}]{hoe15}
{Hoekstra}, H., {Herbonnet}, R., {Muzzin}, A., {Babul}, A., {Mahdavi}, A.,
  {Viola}, M., \& {Cacciato}, M. 2015, \mnras, 449, 685, 1502.01883

\bibitem[{Itoh {et~al.}(1998)Itoh, Kohyama, \& Nozawa}]{relat_corr}
Itoh, N., Kohyama, Y., \& Nozawa, S. 1998, \apj, 502, 7, astro-ph/9712289

\bibitem[{{Kompaneets}(1956)}]{Kompaneets}
{Kompaneets}, A. 1956, Zh.E.F.T., 31, 876

\bibitem[{{Korngut} {et~al.}(2011){Korngut}, {Dicker}, {Reese}, {Mason},
  {Devlin}, {Mroczkowski}, {Sarazin}, {Sun}, \& {Sievers}}]{kor11}
{Korngut}, P.~M. {et~al.} 2011, \apj, 734, 10, 1010.5494

\bibitem[{Lee \& Suto(2004)}]{NonparamSimu2}
Lee, J., \& Suto, Y. 2004, \apj, 601, 599L

\bibitem[{Mayet {et~al.}(2016)}]{NIKA2LP}
Mayet, F., {et~al.} 2016, 1602.07941

\bibitem[{{Mazzotta} {et~al.}(2004){Mazzotta}, {Rasia}, {Moscardini}, \&
  {Tormen}}]{maz04}
{Mazzotta}, P., {Rasia}, E., {Moscardini}, L., \& {Tormen}, G. 2004, \mnras,
  354, 10, arXiv:astro-ph/0409618

\bibitem[{Monfardini {et~al.}(2011)}]{NIKA_cam1}
Monfardini, A., {et~al.} 2011, \apjs, 194, 24, 1102.0870

\bibitem[{{Moreno}(2010)}]{model_moreno}
{Moreno}, R. 2010, {Neptune and Uranus planetary brightness temperature
  tabulation}, Tech. rep., ESA Herschel Science Center, available from
  ftp://ftp.sciops.esa.int/pub/hsc-calibration/PlanetaryModels/ESA2

\bibitem[{Muchovej {et~al.}(2007)}]{SZA}
Muchovej, S., {et~al.} 2007, \apj, 663, 708, astro-ph/0610115

\bibitem[{{Nagai} {et~al.}(2007){Nagai}, {Kravtsov}, \& {Vikhlinin}}]{nag07}
{Nagai}, D., {Kravtsov}, A.~V., \& {Vikhlinin}, A. 2007, \apj, 668, 1,
  arXiv:astro-ph/0703661

\bibitem[{Nagai {et~al.}(2007)Nagai, Vikhlinin, \& Kravtsov}]{gnfw_prof}
Nagai, D., Vikhlinin, A., \& Kravtsov, A.~V. 2007, \apj, 655, 98,
  astro-ph/0609247

\bibitem[{Nord {et~al.}(2009)Nord, Basu, Pacaud, {et~al.}}]{NonparamNord}
Nord, M., Basu, K., Pacaud, F., {et~al.} 2009, \aap, 506, 623

\bibitem[{Perrott {et~al.}(2015)}]{AMI_followup}
Perrott, Y.~C., {et~al.} 2015, \aap, 580, A95, 1405.5013

\bibitem[{{Plagge} {et~al.}(2010){Plagge}, {Benson}, {Ade}, {Aird}, {Bleem},
  {Carlstrom}, {Chang}, {Cho}, {Crawford}, {Crites}, {de Haan}, {Dobbs},
  {George}, {Hall}, {Halverson}, {Holder}, {Holzapfel}, {Hrubes}, {Joy},
  {Keisler}, {Knox}, {Lee}, {Leitch}, {Lueker}, {Marrone}, {McMahon}, {Mehl},
  {Meyer}, {Mohr}, {Montroy}, {Padin}, {Pryke}, {Reichardt}, {Ruhl},
  {Schaffer}, {Shaw}, {Shirokoff}, {Spieler}, {Stalder}, {Staniszewski},
  {Stark}, {Vanderlinde}, {Vieira}, {Williamson}, \& {Zahn}}]{pla10}
{Plagge}, T. {et~al.} 2010, \apj, 716, 1118, 0911.2444

\bibitem[{Plagge {et~al.}(2013)}]{Carmasz}
Plagge, T.~J., {et~al.} 2013, \apj, 770, 112, 1203.2175

\bibitem[{{Planck Collaboration} {et~al.}(2011)}]{XMM_followup}
{Planck Collaboration}, {et~al.} 2011, \aap, 536, A9, 1101.2025

\bibitem[{{Planck Collaboration} {et~al.}(2013)}]{Planck_pressure_prof}
------. 2013, \aap, 550, A131, 1207.4061

\bibitem[{{Planck Collaboration} {et~al.}(2014{\natexlab{a}})}]{Planck_calib}
------. 2014{\natexlab{a}}, \aap, 571, A8, 1303.5069

\bibitem[{{Planck Collaboration} {et~al.}(2014{\natexlab{b}})}]{Planck_cata}
------. 2014{\natexlab{b}}, \aap, 571, A29, 1303.5089

\bibitem[{{Planck Collaboration} {et~al.}(2015{\natexlab{a}})}]{PSZ1_updated}
------. 2015{\natexlab{a}}, \aap, 581, A14, 1502.00543

\bibitem[{{Planck Collaboration} {et~al.}(2015{\natexlab{b}})}]{Planck_param}
------. 2015{\natexlab{b}}, 1502.01589

\bibitem[{{Planck Collaboration} {et~al.}(2015{\natexlab{c}})}]{ymap_planck}
------. 2015{\natexlab{c}}, 1502.01596

\bibitem[{{Planck Collaboration}
  {et~al.}(2015{\natexlab{d}})}]{Planck_SZ_cosmo2015}
------. 2015{\natexlab{d}}, 1502.01597

\bibitem[{{Planck Collaboration} {et~al.}(2015{\natexlab{e}})}]{Planck_cata2}
------. 2015{\natexlab{e}}, 1502.01598

\bibitem[{Pratt {et~al.}(2010)Pratt, Arnaud, Piffaretti, Boehringer, Ponman,
  Croston, Voit, Borgani, \& Bower}]{entropy_REXCESS}
Pratt, G.~W. {et~al.} 2010, \aap, 511, A85, 0909.3776

\bibitem[{{Pratt} {et~al.}(2009){Pratt}, {Croston}, {Arnaud}, \&
  {B{\"o}hringer}}]{pra09}
{Pratt}, G.~W., {Croston}, J.~H., {Arnaud}, M., \& {B{\"o}hringer}, H. 2009,
  \aap, 498, 361, 0809.3784

\bibitem[{Puchwein {et~al.}(2008)Puchwein, Sijacki, \&
  Springel}]{NonparamSimu1}
Puchwein, E., Sijacki, D., \& Springel, V. 2008, \apj, 687, L53

\bibitem[{{Riess} {et~al.}(2007){Riess}, {Strolger}, {Casertano}, {Ferguson},
  {Mobasher}, {Gold}, {Challis}, {Filippenko}, {Jha}, {Li}, {Tonry}, {Foley},
  {Kirshner}, {Dickinson}, {MacDonald}, {Eisenstein}, {Livio}, {Younger}, {Xu},
  {Dahl{\'e}n}, \& {Stern}}]{rie07}
{Riess}, A.~G. {et~al.} 2007, \apj, 659, 98, astro-ph/0611572

\bibitem[{Sayers {et~al.}(2012)Sayers, Czakon, Bridge, Golwala, Koch, Lin,
  Molnar, \& Umetsu}]{Bolocamsz}
Sayers, J., Czakon, N.~G., Bridge, C., Golwala, S.~R., Koch, P.~M., Lin, K.-Y.,
  Molnar, S.~M., \& Umetsu, K. 2012, \apj, 749, L15, 1112.5151

\bibitem[{Sayers {et~al.}(2013)}]{SayersPointSource}
Sayers, J., {et~al.} 2013, \apj, 768, 177, 1211.1632

\bibitem[{Schwan {et~al.}(2011)}]{Apexsz}
Schwan, D., {et~al.} 2011, Rev. Sci. Instrum., 82, 091301, 1008.0342

\bibitem[{{Sehgal} {et~al.}(2011){Sehgal}, {Trac}, {Acquaviva}, {Ade},
  {Aguirre}, {Amiri}, {Appel}, {Barrientos}, {Battistelli}, {Bond}, {Brown},
  {Burger}, {Chervenak}, {Das}, {Devlin}, {Dicker}, {Bertrand Doriese},
  {Dunkley}, {D{\"u}nner}, {Essinger-Hileman}, {Fisher}, {Fowler}, {Hajian},
  {Halpern}, {Hasselfield}, {Hern{\'a}ndez-Monteagudo}, {Hilton}, {Hilton},
  {Hincks}, {Hlozek}, {Holtz}, {Huffenberger}, {Hughes}, {Hughes}, {Infante},
  {Irwin}, {Jones}, {Baptiste Juin}, {Klein}, {Kosowsky}, {Lau}, {Limon},
  {Lin}, {Lupton}, {Marriage}, {Marsden}, {Martocci}, {Mauskopf}, {Menanteau},
  {Moodley}, {Moseley}, {Netterfield}, {Niemack}, {Nolta}, {Page}, {Parker},
  {Partridge}, {Reid}, {Sherwin}, {Sievers}, {Spergel}, {Staggs}, {Swetz},
  {Switzer}, {Thornton}, {Tucker}, {Warne}, {Wollack}, \& {Zhao}}]{seh11}
{Sehgal}, N. {et~al.} 2011, \apj, 732, 44, 1010.1025

\bibitem[{Sembolini {et~al.}(2014)Sembolini, Petris, Yepes, Foschi, Lamagna, \&
  Gottlöber}]{MUSIC_scaling}
Sembolini, F., Petris, M.~D., Yepes, G., Foschi, E., Lamagna, L., \&
  Gottlöber, S. 2014, MNRAS, 440, 3520, 1309.5387

\bibitem[{{Sif{\'o}n} {et~al.}(2016){Sif{\'o}n}, {Battaglia}, {Hasselfield},
  {Menanteau}, {Barrientos}, {Bond}, {Crichton}, {Devlin}, {D{\"u}nner},
  {Hilton}, {Hincks}, {Hlozek}, {Huffenberger}, {Hughes}, {Infante},
  {Kosowsky}, {Marsden}, {Marriage}, {Moodley}, {Niemack}, {Page}, {Spergel},
  {Staggs}, {Trac}, \& {Wollack}}]{sif16}
{Sif{\'o}n}, C. {et~al.} 2016, \mnras, 1512.00910

\bibitem[{{Staguhn} {et~al.}(2008){Staguhn}, {Allen}, {Benford}, {Sharp},
  {Ames}, {Arendt}, {Chuss}, {Dwek}, {Kovacs}, {Maher}, {Marx}, {Miller},
  {Moseley}, {Navarro}, {Sievers}, {Voellmer}, \& {Wollack}}]{sta08}
{Staguhn}, J. {et~al.} 2008, Journal of Low Temperature Physics, 151, 709

\bibitem[{Sunyaev \& Zeldovich(1972)}]{sun72}
Sunyaev, R.~A., \& Zeldovich, Y.~B. 1972, Comments on Astrophysics and Space
  Physics, 4, 173

\bibitem[{Sunyaev \& Zeldovich(1980)}]{SZ_effect}
Sunyaev, R.~A., \& Zeldovich, {\relax Ya}.~B. 1980, \araa, 18, 537

\bibitem[{{Umetsu} {et~al.}(2014){Umetsu}, {Medezinski}, {Nonino}, {Merten},
  {Postman}, {Meneghetti}, {Donahue}, {Czakon}, {Molino}, {Seitz}, {Gruen},
  {Lemze}, {Balestra}, {Ben{\'{\i}}tez}, {Biviano}, {Broadhurst}, {Ford},
  {Grillo}, {Koekemoer}, {Melchior}, {Mercurio}, {Moustakas}, {Rosati}, \&
  {Zitrin}}]{ume14}
{Umetsu}, K. {et~al.} 2014, \apj, 795, 163, 1404.1375

\bibitem[{Vikhlinin(2006)}]{gamma_val}
Vikhlinin, A. 2006, \apj, 640, 710, astro-ph/0504098

\bibitem[{Vikhlinin {et~al.}(2006)Vikhlinin, Kravtsov, Forman, Jones,
  Markevitch, Murray, \& Van~Speybroeck}]{SVM_prof}
Vikhlinin, A., Kravtsov, A., Forman, W., Jones, C., Markevitch, M., Murray,
  S.~S., \& Van~Speybroeck, L. 2006, \apj, 640, 691, astro-ph/0507092

\bibitem[{Voit(2005)}]{Voit_cosmo}
Voit, G.~M. 2005, Rev. Mod. Phys., 77, 207, astro-ph/0410173

\bibitem[{Voit {et~al.}(2005)Voit, Kay, \& Bryan}]{entropy_base}
Voit, G.~M., Kay, S.~T., \& Bryan, G.~L. 2005, MNRAS, 364, 909,
  astro-ph/0511252

\bibitem[{Witzel(1979)}]{spectral_index}
Witzel, A.; Pauliny-Toth, I. I. K. N. U. S.~J. 1979, AJ, 84, 942

\bibitem[{Yu {et~al.}(2015)Yu, Nelson, \& Nagai}]{mergers_bias}
Yu, L., Nelson, K., \& Nagai, D. 2015, \apj, 807, 12, 1501.00317

\end{thebibliography}
\end{document}